\documentclass{elsarticle}

\usepackage{multirow}
\usepackage{rotating}
\usepackage[colorlinks=true, allcolors=blue]{hyperref}
\usepackage{textcomp}
\usepackage{float}
\usepackage[caption = false]{subfig}
\usepackage{epsfig,amssymb,amsmath}
\usepackage[utf8]{inputenc}
\usepackage{graphicx}
\usepackage{amssymb,amsmath,epsfig,url,mathrsfs} 
\usepackage{algorithmic}
\usepackage{algorithm}
\usepackage[usenames, dvipsnames]{color}
\usepackage[framemethod=TikZ]{mdframed}
\usepackage{xcolor}
\usepackage{colortbl}
\usepackage[colorinlistoftodos]{todonotes}
\usepackage{enumitem}
\usepackage{url}

\usepackage{tikz}
\usetikzlibrary{shapes,snakes}

\renewcommand{\vec}[1]{\boldsymbol{\mathrm{#1}}}
\newcommand{\mtx}[1]{\boldsymbol{\mathrm{#1}}}
\newcommand{\transp}{\ensuremath{^\mathsf{T}}}

\newcommand\E{\mathbb{E}}

\newcommand{\diag}[1]{\ensuremath{ \mathsf{diag} \left(#1\right)}}

\renewcommand{\vec}{\boldsymbol} % Vector

\usepackage{amsthm}

\newcommand{\revcolor}{blue} 
 
\makeatletter
\newcommand{\mathleft}{\@fleqntrue\@mathmargin0pt}
\newcommand{\mathcenter}{\@fleqnfalse}
\makeatother
 
\theoremstyle{definition}

\mdfdefinestyle{MyFrame}{%
    linecolor=black,
    outerlinewidth=.5pt,
    roundcorner=5pt,
    innertopmargin=\baselineskip,
    innerbottommargin=\baselineskip,
    innerrightmargin=5pt,
    innerleftmargin=0pt,
    backgroundcolor=gray!10!white}
\definecolor{codegreen}{rgb}{0,0.6,0}
\definecolor{codegray}{rgb}{0.5,0.5,0.5}
\definecolor{codepurple}{rgb}{0.58,0,0.82}
\definecolor{backcolour}{rgb}{0.95,0.95,0.92}

\journal{Computer Speech and Language}

\begin{document}

\begin{frontmatter}

\title{Deep Generative Variational Autoencoding for Replay Spoof Detection in Automatic Speaker Verification\footnote{\textcopyright$2020$. This manuscript version is made available under the CC-BY-NC-ND 4.0 license: \url{http://creativecommons.org/licenses/by-nc-nd/4.0/}}}
\author{Bhusan Chettri$^{1,2}$, Tomi Kinnunen$^1$, Emmanouil Benetos$^2$}
\address{$^1$School of Computing, University of Eastern Finland, FI-80101, Joensuu, Finland \\ $^2$School of EECS, Queen Mary University of London, United Kingdom}

\begin{abstract}
\emph{Automatic speaker verification} (ASV) systems are highly vulnerable to presentation attacks, also called spoofing attacks. \emph{Replay} is among the simplest attacks to mount --- yet difficult to detect reliably. The generalization failure of spoofing countermeasures (CMs) has driven the community to study various alternative deep learning CMs. The majority of them are \emph{supervised} approaches that learn a human-spoof discriminator. In this paper, we advocate a different, \emph{deep generative} approach that leverages from powerful \emph{unsupervised} manifold learning in classification. The potential benefits include the possibility to sample new data, and to obtain insights to the latent features of genuine and spoofed speech. To this end, we propose to use \emph{variational autoencoders} (VAEs) as an alternative backend for replay attack detection, via three alternative models that differ in their class-conditioning. The first one, similar to the use of Gaussian mixture models (GMMs) in spoof detection, is to train independently two VAEs --- one for each class. The second one is to train a single conditional model (C-VAE) by injecting a one-hot class label vector to the encoder and decoder networks. Our final proposal integrates an \emph{auxiliary classifier} to guide the learning of the latent space. Our experimental results using constant-Q cepstral coefficient (CQCC) features on the ASVspoof 2017 and 2019 physical access subtask datasets indicate that the C-VAE offers substantial improvement in comparison to training two separate VAEs for each class. On the 2019 dataset, the C-VAE outperforms the VAE and the baseline GMM by an absolute $9$ - $10$\% in both equal error rate (EER) and tandem detection cost function (t-DCF) metrics. Finally, we propose VAE residuals --- the absolute difference of the original input and the reconstruction as features for spoofing detection. The proposed frontend approach augmented with a convolutional neural network classifier demonstrated substantial improvement over the VAE backend use case.

\end{abstract}


\end{frontmatter}

\section{Introduction}

Voice biometric systems use \emph{automatic speaker verification} (ASV) \cite{reynolds_SC1995} technologies for user authentication. Even if it is among the most convenient means of biometric authentication, the robustness and security of ASV in the face of \emph{spoofing attacks} (or \emph{presentation attacks}) is of growing concern, and is now well acknowledged by the community \cite{sahid_PAD_book}. A spoofing attack involves illegitimate access to personal data of a targeted user.

The ISO/IEC 30107-1 standard \cite{iso_spoofing_standards} identifies nine different points a biometric system could be attacked from (see Fig.~1 in \cite{iso_spoofing_standards}). The first two attack points are of specific interest as they are particularly vulnerable in terms of enabling an adversary to inject spoofed biometric data. These involve: 1) presentation attack at the sensor (microphone in case of ASV); and 2) modifying biometric samples to bypass the sensor. These two modes of attack are respectively known as \emph{physical access} (PA) and \emph{logical access} (LA) attacks in the context of ASV. Artificial speech generated through \emph{text-to-speech} (TTS) \cite{Masuko99onthe} and modified speech generated through \emph{voice conversion} (VC) \cite{PellomH99} can be used to trigger LA attacks. Playing back pre-recorded speech samples (\emph{replay} \cite{wu_APSIPA2014}) and \emph{impersonation} \cite{lau_2004} are, in turn, examples of PA spoofing attacks. Therefore, spoofing countermeasures are of paramount interest to protect ASV systems from such attacks. In this study, a \emph{countermeasure} (CM) is understood as a binary classifier designed to discriminate real human speech or bonafide samples from spoofed ones in a speaker-independent setting.
  
\begin{figure}[t!] %[t]
	\centering  
	\includegraphics[width=\linewidth]{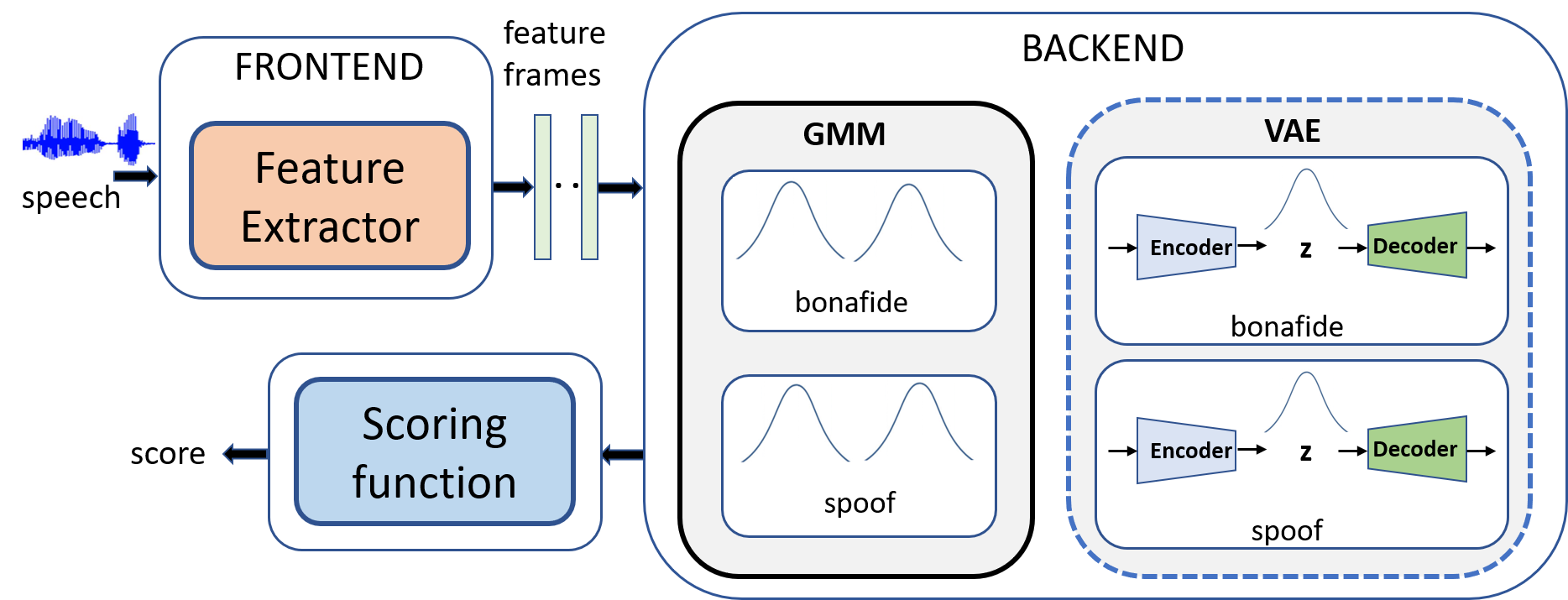}
	%\vspace{-2.7cm}
	\caption{An automatic spoofing detection pipeline using a generative model backend. We explore VAEs as an alternative to a GMM backend. In either case, two generative models are trained to learn the distribution of the bonafide and spoof class data. During inference, for a given test utterance, the scoring function computes the log-likelihood difference between the two generative models as a score. See Section~\ref{background} for details on the methodology.}
	\label{countermeasure}
\end{figure}

Like any traditional machine learning classifier, a spoofing countermeasure (Fig.~\ref{countermeasure}) typically consists of a frontend and a backend module. The key function of the front-end is to transform the raw acoustic waveform to a sequence of \emph{short-term feature vectors}. These short-term feature vectors are then used to derive either intermediate recording-level features (such as \emph{i-vectors} \cite{khoury_ivec_antispoofing,novoseloy_IS2015} or \emph{x-vectors} \cite{jennifer2019challenge}) or statistical models, such as \emph{Gaussian mixture models} (GMMs) \cite{patel_IS2015} to be used for bonafide or spoof class modeling. In contrast to these approaches that require a certain level of handcrafting especially in the frontend, modern \emph{deep-learning} based countermeasures are often trained using either raw-audio waveforms \cite{hannah_raw_spoofing_detection,dinkel2017} or an intermediate high-dimensional time-frequency representation --- often the power spectrogram \cite{zhang_JSTSP2017,galina_IS2017}. In these approaches, the notions of frontend and backend are less clearly distinguished.

In classic automatic speech recognition (ASR) systems and many other speech applications, prior knowledge of speech acoustics and speech perception has guided the design of some successful feature extraction techniques, \emph{mel frequency cepstral coefficients} (MFCCs) \cite{mfccReference} being a representative example. Similar \emph{a priori} characterization of acoustic cues that are relevant for spoofing attack detection, however, is challenging; this is because many attacks are \emph{unseen}, and since the human auditory system has its limits --- it is not designed to detect spoofed speech and may therefore be a poor guide in feature crafting. This motivates the study of data-driven approaches that learn automatically the relevant representations for spoofing detection. In this study, we primarily focus on deep learning based CMs.

Both discriminative models (such as \emph{support vector machines} (SVMs), \emph{deep neural networks} (DNNs) \cite{secondbestsystem_2017challenge,zhang_JSTSP2017}) and generative models (such as GMMs) \cite{patel_IS2015,galina_IS2017}, have extensively been used as backends for spoofing detection. The former directly optimize the class decision boundary while the latter model the data generation process within each of the classes, with the decision boundary being implied indirectly. Both approaches can be used for classification but only the generative approach can be used to sample new data points. We focus on generative modeling as it allows to interpret the generated samples to gain insights about our modeling problem, or to ``debug" the deep learning models and illustrate what the model has learned from the data to make decisions. Further, they can be used for data augmentation which is challenging using purely discriminative approaches.

GMMs have empirically been demonstrated to be competitive in both the ASVspoof 2015 and ASVspoof 2017 challenges \cite{patel_IS2015,galina_IS2017}. While \cite{patel_IS2015} use hand-crafted features for synthetic speech detection, \cite{galina_IS2017} used deep features to train GMM backends. A known problem with GMMs, however, is that use of high-dimensional (short-term) features often leads to numerical problems due to singular covariance matrices. Even if off-the-shelf dimensionality reduction methods such as \emph{principal component analysis} (PCA) \cite{pca_Jolliffe} or \emph{linear discriminant analysis} (LDA) \cite{lda_tharwat} prior to GMM modeling may help, they are not jointly optimized with the GMM. Is there an alternative way to learn a generative model that can handle high-dimensional inputs natively? 

Three generative models that have been demonstrated to produce excellent results in different applications include \emph{generative adversarial networks} (GANs) \cite{goodfellow_GANs}, \emph{variational autoencoders} (VAEs) \cite{kingma2013autoencoding} and \emph{autoregressive} models (for example, \emph{WaveNet} \cite{wavenet}). A VAE is a deep generative probabilistic model that consists of an \emph{encoder} and a \emph{decoder} network. The encoder (inference network) transforms the input data $\vec{x}$ into a low-dimensional \emph{latent} representation, $\vec{z}$, by learning a conditional probability distribution $q_{\vec{\phi}}(\vec{z} |\vec{x})$. The decoder (or generator) network, on the other hand, learns to reconstruct the original data from the low-dimensional latent vector $\vec{z}$ by learning the conditional probability distribution $p_{\theta}(\vec{x} | \vec{z})$. GANs, on the other hand, do not have an encoder/recognition network. Instead, they consist of a generator and a discriminator network. The generator takes as input a random vector and aims to generate fake data that is as close to the real data $\vec{x}$, and the discriminator network aim is to discriminate between the real and the fake data. Autoregressive models, on the other hand, do not use any latent variables. 

Both GANs and VAEs have demonstrated promising results in computer vision \cite{vae_images_label_caption, gulrajani2016pixelvae, vae_image_forecasting}, video generation \cite{mocogan} and natural language processing tasks \cite{gans_NLP}. VAEs have recently been studied for modeling and generation of speech signals \cite{vae_speechModeling, hsu_speech_vae_NIPS, hsu_speech_vae_Interspeech}, and synthesizing music sounds in \cite{vae_music_synthesis}. They have also been used for speech enhancement \cite{simon_speech_enh_ICASSP2019, vae_multichannel_vc_dataset} and feature learning for ASR \cite{feature_learning_asr_vae,asr_applications_vae}. Recent studies in ASV have studied the use of VAEs in \emph{data augmentation} \cite{dataaugmentation_vae_asv}, \emph{regularisation} \cite{regularisation_vae_asv} and \emph{domain adaptation} \cite{adaptation_vae_asv} for deep speaker embeddings (x-vectors). In TTS, VAEs have been recently used to learn speaking style in an end-to-end setting \cite{zhang_vae_tts}. Recent work in \cite{yexin2019challenge} uses VAEs for extracting low-dimensional features and trains a separate classifier on these features for spoofing detection. However, as far as the authors are aware, the  application of VAEs as a backend classifier for spoofing attack detection in ASV remains an unexplored avenue.

In this work, we focus on deep probabilistic VAEs as a backend for spoofing detection. Figure \ref{countermeasure} illustrates our idea. While VAEs have both inference and generator networks, GANs do not have an inference network and autoregressive models do not use latent variables.
This motivates our focus on VAEs over other deep generative models, as it suits more naturally our task. %formulation illustrated in Figure \ref{countermeasure}. 
The reconstruction quality of VAEs tends to be inferior to that obtained by GANs \cite{Huang_vae_gan}, but for classification tasks, obtaining a perfect reconstruction is \emph{not} the main priority. A key challenge, instead, is how to train VAEs to not only preserve reasonable reconstruction but to allow to retain discriminative information in the latent space. To address this, VAEs are often trained with additional constraints. For example, by conditioning the encoder and decoder with additional input --- so called \emph{conditional} VAE (C-VAE) \cite{conditional_vae}; or by augmenting an \emph{auxiliary classifier} either to the latent space \cite{speech_separation_vae_auxiliary_classifier_ICASSP} or to the output of the decoder network \cite{auxiliary_classifier_vae_vc}. As there is no \emph{de facto} standard on this, we aim to fill this knowledge gap in the domain of audio replay detection. In specific, we present a feasibility study of various alternative VAEs for replay spoofing attack detection. 
We summarize the contributions of this work as follows:
\begin{itemize}
	\item While deep generative models, VAEs in particular, have been studied in many other domains, their application in audio spoofing detection remains less explored to date. We study the potential of deep generative VAEs as a backend classifier for spoofing detection. To the best of our knowledge, this is the first work in this direction.
	
	\item We describe practical challenges in training a VAE model for spoofing detection applications and discuss approaches that can help overcome them, which could serve as potential guidelines for others.
	
	\item Along with a ``naive"\footnote{We use \emph{naive} VAE to refer the standard (vanilla) VAE \cite{kingma2013autoencoding} trained without any class labels. Information about the class is included by independently training one VAE per class.} VAE we also study conditional VAEs (C-VAEs) \cite{conditional_vae}. The C-VAE uses class labels as an additional conditional input during training and inference. Since we pass class labels in C-VAE, we use a single model to represent both classes unlike naive VAE where we train two separate models, one each for bonafide and spoof class. For the text-dependent setting of ASVspoof 2017 data, we further address conditioning using a combination of the class and passphrase labels.
	
	\item Inspired by \cite{speech_separation_vae_auxiliary_classifier_ICASSP, auxiliary_classifier_vae_vc}, we introduce an \emph{auxiliary classifier} into our VAE modeling framework and study how this helps the latent space\footnote{A latent space is a probability distribution that defines the observed-data generation process and is characterised by means and variances of the encoder network.} to capture discriminative information without sacrificing much during reconstruction. We experiment adding the classifier model on top of the latent space and at the end of the decoder (Section \ref{different_vae_models}).
	
	\item While our primary focus is in using VAEs as a back-end, we also address their potential in feature extraction. In particular, we subtract a VAE-modeled spectrogram from the original spectrogram so as to de-emphasize the importance of salient speech features (hypothesized to be less relevant in spoofing attack detection) and focus on the residual details. We train a convolutive neural network classifier using these residual features.
	
\end{itemize}

\begin{sidewaystable}
	\centering
	\caption{Summary of deep learning architectures/methods studied in the context of spoofing detection in ASV. Disc: discriminative, Gen: generative. D1: ASVspoof $2015$, D2: BTAS $2016$, D3.1: ASVspoof $2017$ v1.0, D3.2: ASVspoof $2017$ v2.0, D4: ASVspoof $2019$ LA, D5: ASVspoof $2019$ PA. Metrics reported are equal error rate, EER, and tandem detection cost function, t-DCF, unless otherwise stated. Reported numbers are for the respective evaluation sets.}
	\scalebox{0.90}{
	\begin{tabular}{cccccccc}
   	\hline    
     \textbf{Modeling} &\textbf{Model/} &\textbf{Input} &\textbf{Purpose} &\textbf{Metric} &\textbf{Dataset} &\textbf{Fusion} &\textbf{anti-spoofing} \\
   	    
   	\textbf{approach} &\textbf{Architecture} &\textbf{features} &\textbf{} &\textbf{[EER/t-DCF]} &\textbf{used} &\textbf{}   &\textbf{reference} \\
   	
   	\hline
   	Disc &CNN &raw-waveform &end-to-end  &$0.157$ &D1 &- &\cite{hannah_raw_spoofing_detection}  \\
   	Disc &CNN+LSTM &raw-waveform &end-to-end &$0.82$ HTER &D2 &- &\cite{dinkel2017}  \\
   	Disc &CNN &spectrogram &classification &$10.6$ &D3.2 &- &\cite{bhusanSLT2018} \\
   	Disc &LCNN \cite{wu2015light} &spectrogram &embedding learning &- &D3.1 &- &\cite{galina_IS2017}   \\
   	Gen  &GMMs &embedding &classification &$7.37$ &D3.1 &- &\cite{galina_IS2017}  \\
   	Disc&CNN &CQCC+HFCC &embedding learning &$11.5$ &D3.1 &feature &\cite{secondbestsystem_2017challenge}  \\
   	
   	Disc &DNN+RNN  &FBanks, MFCCs &embedding learning & - &D1 &-  &\cite{yanmin2016} \\
   	Disc &SVM, LDA &embedding &classification &$1.1$ &D1 &score &\cite{yanmin2016} \\ 	               
   	Disc &CNN &  &embedding learning &- &D3.1 &- &\cite{kaavya_IS2018}  \\
   	Gen &GMM &embedding &classification &$6.4$ &D3.1 &- &\cite{kaavya_IS2018}  \\	   	       
   	
   	Disc &Bayesian DNN, LCNN &spectrogram &classification &[$0.88$/$0.0219$] &D5 &score & \cite{bayesian2019challenge}  \\
   	
   	Disc &VGG, SincNet, LCNN &spectrogram, CQT &classification &[$1.51$/$0.0372$] &D5 &score & \cite{but2019challenge}  \\
   	& & &  &[$8.01$/$0.2080$] &D4 &score & \cite{but2019challenge} \\         
   	
   	Disc &LCNN, RNN &spectrogram &embedding learning &- &- &- & \cite{alejandro2019challenge}  \\
   	Gen &PLDA &embedding &classification &$6.08$ &D3.1 &- & \cite{alejandro2019challenge}  \\
   	Disc &LDA &embedding &classification &[$6.28$ /$0.1523$] &D4 &- & \cite{alejandro2019challenge}  \\
   	Gen &PLDA &embedding &classification &[$2.23$ /$0.0614$] &D5 &- & \cite{alejandro2019challenge}  \\
   	Disc &TDNN, LCNN, ResNet &CQCCs, LFCCs &embedding learning &[$9.08$ /$0.1791$] &D4 &score &\cite{chang2019challenge}     \\
   	
   	Disc &ResNet \cite{he2015deep} &STFT, group delay gram &classification &[$0.66$ /$0.0168$] &D5 &score & \cite{weicheng2019challenge}  \\

   	Disc &TDNN  &MFCC, CQCC, spectrogram  &multitasking, classification &$7.94$ &D3.2 &score &\cite{rongjin2019challenge} \\
   	&  &  & &[$7.63$ /$0.2129$] &D4 &score &\cite{rongjin2019challenge} \\
   	&  &  & &[$0.96$ /$0.0266$] &D5 &score &\cite{rongjin2019challenge} \\   	
   	Disc &ResNet &MFCCs, CQCCs &classification &$13.30$ &D3.1 &score &\cite{fourthbestsystem_2017challenge}  \\
   	
   	Disc &ResNet, SeNet &CQCC, spectrogram &classification &[$0.59$ /$0.016$] &D5 &score &\cite{assert2019challenge}  \\
   	& & & &[$6.7$ /$0.155$] &D4 &score &\cite{assert2019challenge}  \\
   	
   	Disc &ResNet &MFCCs, CQCC, spectrogram &classification &[$6.02$ /$0.1569$] &D4 &score &\cite{moustafa2019challenge}  \\
   	& & & &[$2.78$ /$0.0693$] &D5 &score &\cite{moustafa2019challenge} \\   	    	 
   	
   	Disc &CNN+GRU &spectrogram &classification &[$2.45$ /$0.0570$] &D5 &score &\cite{jee2019challenge}  \\   	
   	&ResNet, LSTMs  &spectrogram &embedding learning &$16.39$ &D3.1 &score &\cite{resnet_dataAugmentation}  \\

   	Gen &GMMs &CQCC, LFCC, MelRP  &classification &$11.43$ &D3.2 &score &\cite{liuICASSP2019}  \\   	
   	Gen &Attention-based ResNet &spectrogram &classification &$8.54$ &D3.2 &score &\cite{laiICASSP2019}  \\   	
   	
   	Gen &VAE &CQT spectrogram &embedding learning &- &D4 &- &\cite{yexin2019challenge}  \\   
   	\hline
    Gen &VAE &CQCC, spectrogram &classification, feature extraction &&D3.2, D5& &\textbf{this study} \\
   	
   	\hline
	\end{tabular}
	\label{literature_summary}}
\end{sidewaystable}

\section{Related work}

\textbf{Traditional methods}. Since the release of benchmark anti-spoofing datasets \cite{wu_IS2015,tomiSummaryPaper,asvspoof2019overview} and evaluation protocols as part of the ongoing ASVspoof challenge series\footnote{http://www.asvspoof.org/}, there has been considerable research on presentation attack detection \cite{sahid_PAD_book}, in particular for TTS, VC, and replay attacks. 
Many anti-spoofing features coupled with a GMM backend have been studied and proposed in the literature. We briefly discuss them here. \emph{Constant Q cepstral coefficients} (CQCCs) \cite{hector_cqcc}, among them, have shown state-of-the-art performance on TTS and VC spoofed speech detection tasks on the ASVspoof 2015 dataset \cite{wu_IS2015}. They have been adapted as baseline features in the recent ASVspoof 2017 and ASVspoof 2019 challenges. Further tweaks on CQCCs have been studied in \cite{rohan_eCQCC} showing some improvement over the standard CQCCs. \emph{Teager energy operator} (TEO) based spoof detection features have been studied in \cite{patel_IS2017}. Speech demodulation features using the TEO and the Hilbert transform have been studied in \cite{madhu_IS2018}. Authors in \cite{buddhi_IS2018} proposed features for spoofing detection by exploiting the long-term temporal envelopes of the subband signal. Spectral centroid based frequency modulation features have been proposed in \cite{tharshini_IS2018}. \cite{saranya_IS2018} proposes the use of decision level feature switching between mel and linear filterbank slope based features, demonstrating promising performance on the ASVspoof 2017 v2.0 dataset. Adaptive filterbank based features for spoofing detection have been proposed in \cite{buddhi2019IS}. Finally, \cite{hardik_IS2018} proposes the use of convolutional \emph{restricted Boltzmann machines} (RBMs) to learn temporal modulation features for spoofing detection.

\textbf{Deep learning methods}. Deep learning-based systems have been proposed either for feature learning \cite{yanmin2016,galina_IS2017,secondbestsystem_2017challenge,kaavya_IS2018} or in an end-to-end setting to model raw audio waveforms directly \cite{dinkel2017,hannah_raw_spoofing_detection}. A number of studies \cite{rongjin2019challenge,chang2019challenge} have also focused on \emph{multi-task learning} for improved generalization by simultaneously learning an auxiliary task. \emph{Transfer learning} and \emph{data augmentation} approaches have been addressed in \cite{chang2019challenge,weicheng2019challenge}. Some of the well known deep architectures from computer vision, including \emph{ResNet} \cite{he2015deep} and \emph{light CNN} \cite{wu2015light} have been adopted for ASV anti-spoofing in \cite{fourthbestsystem_2017challenge,resnet_dataAugmentation, assert2019challenge, moustafa2019challenge, jee2019challenge} and \cite{galina_IS2017, bayesian2019challenge, but2019challenge, alejandro2019challenge}, respectively, demonstrating promising performance on the ASVspoof challenge datasets. The recently proposed \emph{SincNet} \cite{ravanelli2018speaker} architecture for speaker recognition was also proposed for spoofing detection in \cite{but2019challenge}. Furthermore, \emph{attention-based models} have been studied in \cite{liuICASSP2019,laiICASSP2019} during the ASVspoof 2019 challenge. It is also worth noting that the best performing models on the ASVspoof challanges used \emph{fusion} approaches, either at the classifier output or the feature level \cite{assert2019challenge, bhusan2019challenge, galina_IS2017}, indicating the challenges in designing a single countermeasure capable of capturing all the variabilities that may appear in wild test conditions in a presentation attack. Please refer to Table~\ref{literature_summary} for details. %a detailed summary of

As Table \ref{literature_summary} summarizes, there is a substantial body of prior work on deep models in ASV anti-spoofing, even if it is hard to pinpoint commonly-adopted or outstanding methods. Nonetheless, the majority of the approaches rely either on discriminative models or classical (shallow) generative models. This leaves much scope for further studies in \emph{deep generative} modeling. Recently, VAEs have been used for embedding learning for spoofing detection \cite{yexin2019challenge}. They trained a bonafide VAE using only the bonafide utterances from the 2019 LA dataset, and use it to extract $32$ dimensional embeddings for both bonafide and spoof utterances. Unlike \cite{yexin2019challenge}, our main focus is on studying VAEs as a backend classifier, described in the next section.

\section[theory-background]{Methodology} \label{background}

This section provides a brief background to the VAE. Besides the original work \cite{kingma2013autoencoding}, the reader is pointed to tutorials such as \cite{Altosaar2019-vae-tutorial} and \cite{Doersch16-VAE-tutorial} for further details on VAEs. We also make a brief note on the connection between VAEs and Gaussian mixture models (GMMs), both of which are generative models involving latent variables \cite{bishop_2006}. 

\subsection{Variational autoencoder (VAE)}

The \emph{variational autoencoder} (VAE) \cite{kingma2013autoencoding} is a \emph{deep generative model} that aims at uncovering the data generation mechanism in the form of a probability distribution. The VAE is an \emph{unsupervised} approach that learns a low-dimensional, nonlinear data \emph{manifold} from training data without class labels. VAEs achieve this by using two separate but jointly trained neural networks, an \emph{encoder} and a \emph{decoder}. The encoder forces the input data through a low-dimensional \emph{latent space} that the decoder uses to reconstruct the input.

Given a $D$-dimensional input $\vec{x} \in \mathbb{R}^D$, the encoder network maps $\vec{x}$ into a latent vector $\vec{z} \in \mathbb{R}^d$ ($d \ll D$). Unlike in a conventional (deterministic) autoencoder, $\vec{z}$ is not a single point; instead, the encoder imposes a \emph{distribution} over the latent variable, $q_{\vec{\phi}}(\vec{z}|\vec{x})$, where $\vec{\phi}$ denotes all the parameters (network weights) of the encoder. The default choice, also in this work, is a Gaussian $q_{\vec{\phi}}(\vec{z}|\vec{x})=\mathcal{N}(\vec{z}|\vec{\mu}_{\vec{\phi}}(\vec{x}), \diag{\mtx{\vec{\sigma}}_{\vec{\phi}}^{2}({\vec{x}})})$, where $\vec{\mu}_{\vec{\phi}}(\vec{x})$ and
$\diag{\mtx{\vec{\sigma}}_{\vec{\phi}}^{2}({\vec{x}})}$ are deterministic functions (the encoder network) that return the mean and variance vector (\emph{i.e.}, diagonal covariance matrix) of the latent space given input $\vec{x}$. 

The decoder network, in turn, takes $\vec{z}$ as input and returns a parameterized probability distribution, which is another Gaussian. The decoder distribution is $p_{\vec{\theta}}(\vec{x}|\vec{z})=\mathcal{N}(\vec{x}|\vec{\mu}_{\vec{\theta}}(\vec{z}), \diag{\mtx{\vec{\sigma}}_{\vec{\theta}}^{2}({\vec{z}})})$, where $\vec{\mu}_{\vec{\phi}}(\vec{z})$ and
$\diag{\mtx{\vec{\sigma}}_{\vec{\theta}}^{2}({\vec{z}})}$ are deterministic functions implemented by the decoder network, and where $\vec{\theta}$ denotes the decoder network parameters. Random observations sampled from the decoder distribution (with fixed $\vec{z}$) should then bear resemblance to the input $\vec{x}$. In the standard VAE, the only sampling that takes place is from the variational posterior distribution of the latent variable. Conceptually, however, it is useful to note that the decoder also produces a distribution of possible outputs, rather a single point estimate, for a given (fixed) $\vec{z}$. These outputs will not be exactly the same as $\vec{x}$ due to the dimensionality reduction to the lower-dimensional $\vec{z}$-space, but each of the individual elements of the $\vec{z}$-space represents some salient, meaningful features necessary for approximating $\vec{x}$.

\subsection{VAE training}
The VAE is trained by maximizing a regularized log-likelihood function. Let $\mathcal{X}=\{\vec{x}_n\}_{n=1}^N$ denote the training set, with $\vec{x}_n\in \mathbb{R}^D$. The training loss for the entire training set $\mathcal{X}$,
    \begin{equation}
        \mathcal{L}(\vec{\theta},\vec{\phi}) = \sum_{n=1}^N \ell_n(\vec{\theta},\vec{\phi}),
    \end{equation}
decomposes to a sum of data-point specific losses. The loss of the $n$th training example is a regularized reconstruction loss:
    \begin{equation}
        \ell_n(\vec{\theta},\vec{\phi}) = 
        \underbrace{-\E_{\vec{z}\sim q_{\vec{\phi}}(\vec{z}|\vec{x}_n)}\Big[\log p_{\vec{\theta}}(\vec{x}_n|\vec{z})\Big]}_{\text{Reconstruction error}} 
        + \underbrace{\text{KL} \big(q_{\vec{\phi}}(\vec{z}|\vec{x}_n)\,\Vert\, p(\vec{z}) \big)}_{\text{Regularizer}},  \label{eq:ELBO}
    \end{equation}
where $\E[\cdot]$ denotes expected value and $\text{KL}(\cdot \Vert \cdot)$ is the \emph{Kullback-Leibler divergence} \cite{CoverThomas2001-elements} -- a measure of difference between two probability distributions. The reconstruction error term demands for an accurate approximation of $\vec{x}$ while the KL term penalizes the deviation of the encoder distribution from a fixed \emph{prior distribution}, $p(\vec{z})$. Note that the prior, taken to be the standard normal, $p(\vec{z})=\mathcal{N}(\vec{z}|\vec{0},\mtx{I})$, is shared across all the training exemplars. It enforces the latent variables $\vec{z}$ to reside in a compatible feature space across the training exemplars. We use stochastic gradient descent to train all our VAE models. More training details provided later in \ref{model_training}.

In practice, to derive a differentiable neural network after sampling $\vec{z}$, VAEs are trained with the aid of the so-called \emph{reparameterization trick} \cite{kingma2013autoencoding}. Thus, sampling $\vec{z}$ from the posterior distribution $q_{\vec{\phi}}(\vec{z}|\vec{x})$ is performed by computing $\vec{z} = \vec{\mu}_{\vec{\phi}}(\vec{x}) + \vec{\sigma}_{\vec{\phi}}({\vec{x}}) \odot \vec{\epsilon}$ where $\vec{\epsilon}$ is a random vector drawn from  $\mathcal{N}(\vec{z}|\vec{0},\mtx{I})$, $\vec{\mu}$ and $\vec{\sigma}$ are the means and variance of the posterior learned during the VAE training, and $\odot$ denotes the element-wise product.

\begin{figure}
\subfloat[\textbf{Naive VAE}. Separate bonafide and spoof VAE models are trained using the respective-class training audio files.]{\includegraphics[width = \linewidth]{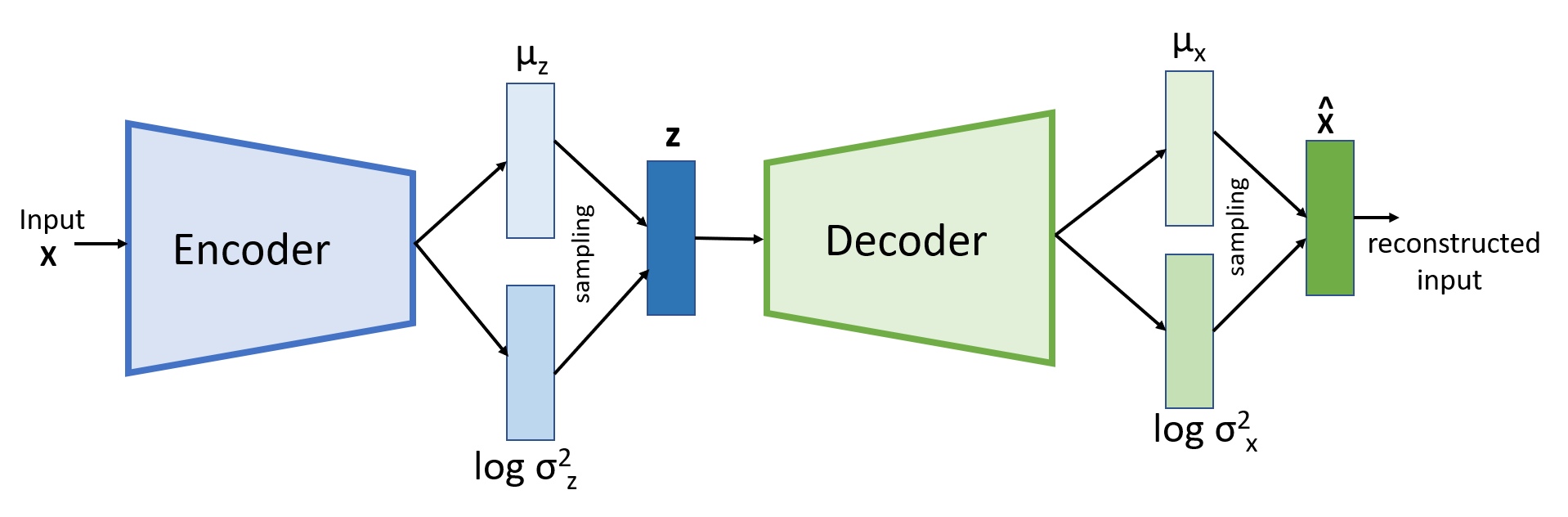}} \\
\subfloat[\textbf{C-VAE}. A single VAE model is trained using the entire training examples but with class-label vectors.]{\includegraphics[width = \linewidth]{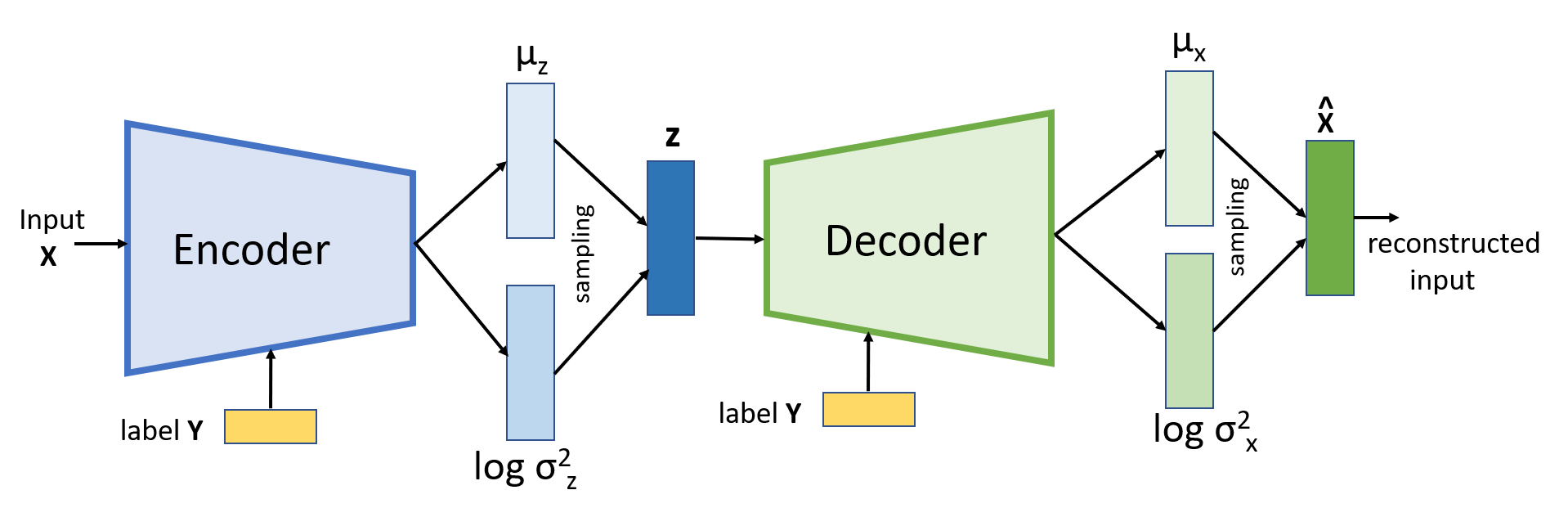}}\\
\subfloat[\textbf{AC-VAE} extends \textbf{C-VAE} by augmenting an auxiliary classifier (AC). We include AC in two alternative settings: (i) AC-VAE$_1$: use latent mean vector $\vec{\mu_{\vec{z}}}$ as its input, or (ii) AC-VAE$_2$: at the end of decoder using reconstruction as its input. These are highlighted with dotted lines. At test time we discard the AC.]{\includegraphics[width = \linewidth]{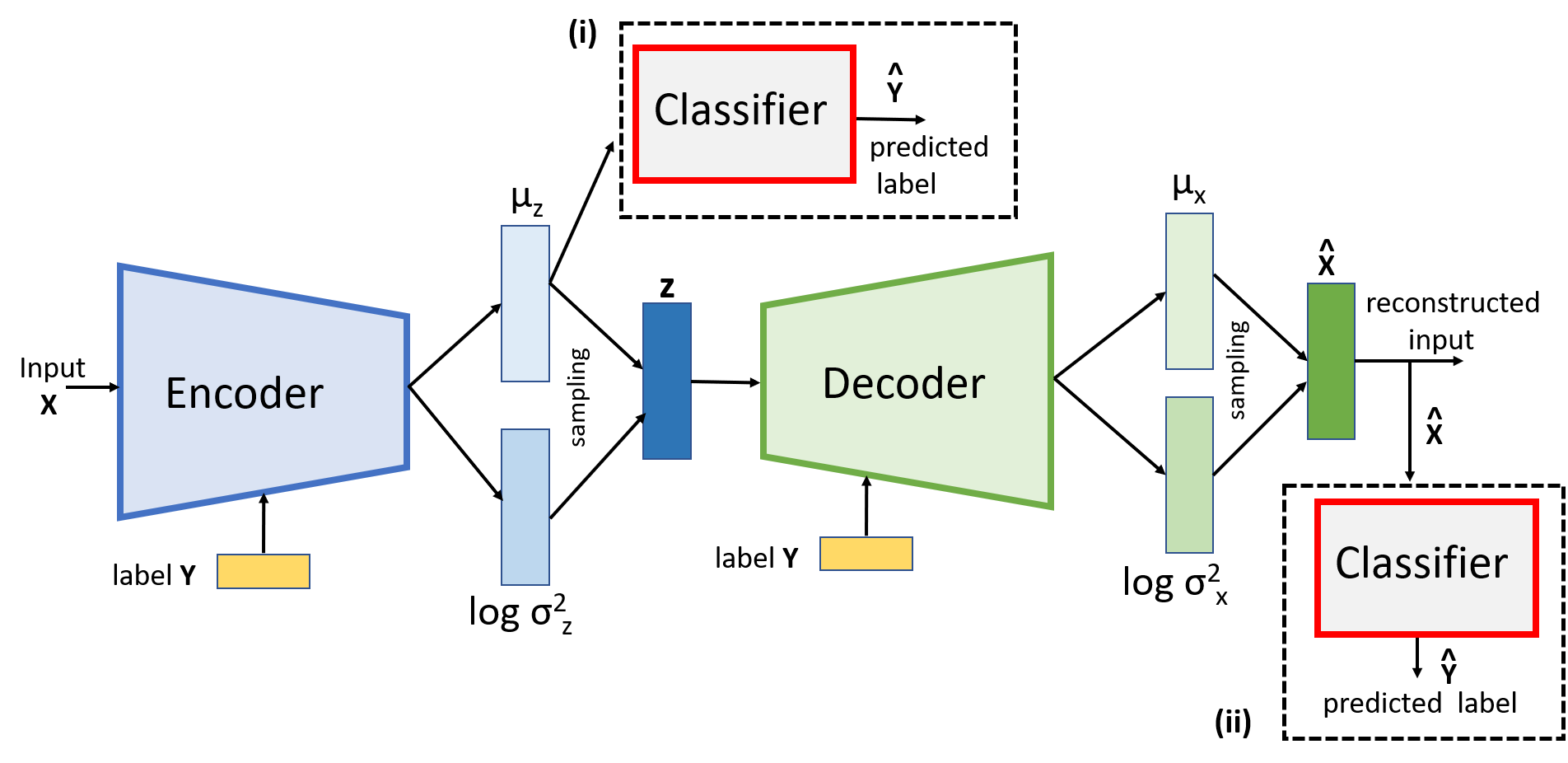}}\\

\caption{Different VAE setups studied in this paper.}
\label{vae_figures}
\end{figure}

\subsection{Conditioning VAEs by class label} \label{different_vae_models}

As said, the VAE is an unsupervised method that learns an encoder-decoder pair, $\vec{\Lambda}=(\vec{\theta},\vec{\phi})$, without requiring class labels. When used for \emph{classification}, rather than data reconstruction, we have to condition VAE training with the class label. Here, we use labels $y_n=1$ (bona fide) and $y_n=0$ (spoof) to indicate whether or not the $n^\text{th}$ training exemplar represents bona fide speech\footnote{We use the vector notation $\vec{y}_n$ to indicate the corresponding \emph{one-hot vector} --- \emph{i.e.}, $\vec{y}_n=(0,1)$ to represent bonafide and $\vec{y}_n=(1,0)$ to represent a spoof sample.}. We consider three alternative approaches to condition VAE training, described as follows.

The first, \textbf{naive} approach, is to train VAEs similarly as GMMs \cite{patel_IS2017,saranya_IS2018,hector_cqcc} --- independently of each other, using the respective training sets $\mathcal{X}_\text{bona}=\{\vec{x}_n | y_n=1\}$ and $\mathcal{X}_\text{spoof}=\{\vec{x}_n | y_n=0\}$. VAEs are trained to optimize the loss function described in \eqref{eq:ELBO}. This yields two VAEs, $\vec{\Lambda}_\text{bona}$ and $\vec{\Lambda}_\text{spoof}$. At the test time, the two VAEs are independently scored using \eqref{eq:ELBO}, and combined by subtracting the spoof score from the bona fide score. The higher the score, the higher the confidence that the test utterance originates from the bonafide class.

Our second approach is to train a single \textbf{conditional VAE} (C-VAE) \cite{conditional_vae} model. In contrast to the naive approach, the C-VAE can learn more complex (\emph{e.g.}, multimodal) distributions by including auxiliary inputs (conditioning variables) to the encoder and/or decoder distributions. In this approach, the label vector $\vec{y}_n$ is used both in training and scoring. Specifically, in our implementation inspired from \cite{cvae_av_speech_synthesis,dataaugmentation_vae_asv}, we augment $\vec{y}_n$ to the output of the last convolutional layer in the encoder network and to the input of the decoder network. Section \ref{vae_architecture} describes our encoder and decoder architectures. The training loss \eqref{eq:ELBO} is now  revised as, 

\begin{equation}
        \ell_n(\vec{\theta},\vec{\phi}) = 
        -\E_{\vec{z}\sim q_{\vec{\phi}}(\vec{z}|\vec{x}_n,\vec{y}_n)}\Big[\log p_{\vec{\theta}}(\vec{x}_n|\vec{z},\vec{y}_n)\Big]
        + \text{KL} \big(q_{\vec{\phi}}(\vec{z}|\vec{x}_n,\vec{y}_n)\,\Vert\, p(\vec{z}\textcolor{\revcolor}{|\vec{y}_n}) \big),    \label{eq:ELBO_cvae}
\end{equation}
where, in practice, we relax the class-conditional prior distribution of the latent variable to be independent of the class, \emph{i.e.} $p(\vec{z}|\vec{y}_n)=p(\vec{z})$  \cite{conditional_vae}. We perform scoring in the same way as for the previous approach: we pass each test exemplar $\vec{x}$ through the C-VAE using genuine and spoof class vectors $\vec{y}_n$, to give two different scores, which are then differenced as before. Note that $\vec{y}_n$ may include any other available useful metadata besides the binary bonafide/spoof class label. In our experiments on the \emph{text-dependent} ASVspoof 2017 corpus consisting of 10 fixed passphrases, we will address the use of class labels and phrase identifiers jointly.

Our third approach is to use an \textbf{auxiliary classifier with a conditional VAE} (AC-VAE) to train a discriminative latent space. We use $r_{\vec{\psi}}(\vec{x})$ to denote the predicted posterior probability of the bonafide class, as given by an auxiliary classifier (AC). And, $\vec{\psi}$ denotes the parameters of AC. Note that the posterior for the spoof class is $1-r_{\vec{\psi}}(\vec{x})$ as there are two classes. Inspired by \cite{adaptation_vae_asv} and \cite{auxiliary_classifier_vae_vc}, we consider two different AC setups. First, following \cite{adaptation_vae_asv}, we use the mean $\vec{\mu_{\vec{z}}}$ as the input to an AC which is a feedforward neural network with a single hidden layer. Second, following \cite{auxiliary_classifier_vae_vc}, we augment a deep-CNN as an AC to the output of the decoder network. Here, we use the CNN architecture from \cite{bhusanSLT2018}. From hereon, we call these two setups as AC-VAE$_1$ and AC-VAE$_2$ respectively. To train the model, we jointly optimise the C-VAE loss \eqref{eq:ELBO_cvae} and the AC loss. In specific, the loss for the $n^\text{th}$ training exemplar is
\begin{equation}
        \ell_n(\vec{\theta},\vec{\phi}, \vec{\psi}) =  \alpha \cdot \ell_n(\vec{\theta},\vec{\phi}) + \beta \cdot \ell_n(\vec{\psi}), \label{aux_classifier_loss}
\end{equation}
where the nonnegative control parameters $\alpha$ and $\beta$ weigh the relative importance of the regularisation terms during training, set by cross-validation, and where $\ell_n(\vec{\psi})$ denotes the binary \emph{cross-entropy loss} for the training exemplar $\vec{x}_n$. It is defined as 

\begin{equation}\label{cross_entropy_loss}
         \ell_n(\vec{\psi}) = -\big(y_n\,\log\,r_{\vec{\psi}}(\vec{x}_n) + (1-y_n)\,\log (1-r_{\vec{\psi}}(\vec{x}_n))\big)
\end{equation} 

\noindent 
Note that setting $\alpha=1$ and $\beta=0$ in \eqref{aux_classifier_loss} gives \eqref{eq:ELBO_cvae} as a special case. At test time we discard the auxiliary classifier and follow the same approach for scoring as in the C-VAE setup discussed earlier. All the three approaches are summarized in Fig. \ref{vae_figures}.

\subsection{Gaussian mixture model (GMM)} \label{gmm_models}

Besides VAEs, our experiments include a standard GMM-based approach. Similar to the VAE, the GMM is also a generative model that includes latent variables. In the case of GMMs, $\vec{x}$ is a short-term feature vector, and $\vec{z}$ is a one-hot vector with $C$ components (the number of Gaussians), indicating which Gaussian was `responsible' for generating $\vec{x}$. Let $\vec{z}_k=(0,0,\dots,1,0,\dots,0)\transp$ be a realization of such one-hot vector where the $k$-th element is 1. The conditional and prior distributions of GMM are:

\begin{equation}
        \begin{aligned}
            p(\vec{x}|\vec{z} & =\vec{z}_k,\vec{\Lambda})=\mathcal{N}(\vec{x}|\vec{\mu}_k,\mtx{\Sigma}_k)\\
            p(\vec{z} & =\vec{z}_k,\vec{\Lambda})=\pi_k,
        \end{aligned}
    \end{equation}
where $\vec{\Lambda}=(\vec{\mu}_k,\mtx{\Sigma}_k,\pi_k)_{k=1}^C$ denotes the GMM parameters (means, covariances and mixing weights). By marginalizing the latent variable out, the log-likelihood function of a GMM is given by:

\begin{equation}\label{eq:gmm-likelihood}
        \log p_{\vec{\Lambda}}(\vec{x}) = \log \sum_{\vec{z}} p(\vec{z})p(\vec{x}|\vec{z}) = \log \sum_{k=1}^C \pi_k\,\mathcal{N}(\vec{x}|\vec{\mu}_k,\mtx{\Sigma}_k),
    \end{equation}
used as a score when comparing test feature $\vec{x}$ against the GMM defined by $\vec{\Lambda}$. At the training stage we train two GMMs (one for bonafide, one for spoof). As maximizing \eqref{eq:gmm-likelihood} does not have a closed-form solution, GMMs are trained with the  \emph{expectation-maximization} (EM) algorithm \cite{bishop_2006,Dempster1977-EM} that iterates between component assignment (a `soft' version of the true 1-hot latent variable $\vec{z}$) and parameter update steps.

\subsection{VAEs and GMMs as latent variable models}

Given the widespread use of GMMs in voice anti-spoofing studies, it is useful to compare and contrast the two. Both are generative approaches, and common to both is the assumption of the data generation process consisting of two consecutive steps:
\begin{mdframed}[style=MyFrame]\label{neutral-role-play}
    \begin{enumerate}
        \item \textbf{First}, one draws a latent variable $\vec{z}_n \sim p_\text{gt}(\vec{z})$ from a \emph{prior distribution}.
        \item \textbf{Second}, given the selected latent variable, one draws the observation from a \emph{conditional distribution}, $\vec{x}_n \sim p_\text{gt}(\vec{x}|\vec{z}_n)$,
    \end{enumerate}
\end{mdframed}    
where the subscript `gt' highlights an assumed underlying `true' data generator whose details are unknown. Both VAEs and GMMs use parametric distributions to approximate $p_\text{gt}(\vec{z})$ and $p_\text{gt}(\vec{x}|\vec{z}_n)$. In terms of the `$\vec{z}$' variable, the main difference between GMMs and VAEs is that in the former it is discrete (categorical) and in the latter it is continuous. As for the second step, in GMMs, one draws the observation from a multivariate Gaussian distribution corresponding to the selected component. In VAEs, one also samples the reconstructed observation from a Gaussian, but the mean and covariance are not selected from an enumerable set --- they are continuous and are predicted by the decoder from a given $\vec{z}$.

Both GMMs and VAEs are trained with the aim of finding model parameters that maximize the training data log-likelihood; common to both is that no closed-form solution for the model parameters exists. The way the two models approach the parameter estimation (learning) problem differs substantially, however. As in any maximum likelihood estimation problem, the training observations are assumed to be i.i.d., enabling the log-likelihood function over the whole training dataset to be written as the sum of log-likelihoods over all the training observations. This holds both for VAE and GMM. Let us use the GMM as an example. For a single observation $\vec{x}$, the log-likelihood function is:
    \begin{equation}\label{eq:gmm-elbo}
        \begin{aligned}
        \log p_{\vec{\Lambda}}(\vec{x}) & = \log \sum_{\vec{z}} p(\vec{x,z}|\vec{\Lambda}) = 
        \sum_{\vec{z}} Q(\vec{z})\frac{p(\vec{x,z}|\vec{\Lambda})}{Q(\vec{z})} =\log\,\, \E_{\vec{z}\sim Q(\vec{z})}\Bigg[\frac{p_{\vec{\Lambda}}(\vec{x,z})}{Q(\vec{z})} \Bigg]\\
        & \geq \E_{\vec{z}\sim Q(\vec{z})}\Bigg[\log \frac{p_{\vec{\Lambda}}(\vec{x,z})}{Q(\vec{z})} \Bigg]
        = \sum_{\vec{z}} Q(\vec{z}) \log \frac{p_{\vec{\Lambda}}(\vec{x,z})}{Q(\vec{z})}
        \end{aligned}
    \end{equation}

\noindent    
where $Q(\vec{z})$ is \emph{any} distribution, and where the inequality in the second line is obtained using \emph{Jensen's inequality} \cite{CoverThomas2001-elements} (using concavity of logarithm). The resulting last expression, known as the \emph{evidence lower bound} (ELBO), serves as a lower bound of the log-likelihood which can be maximized more easily. The well-known EM algorithm \cite{Dempster1977-EM} is an alternating maximization approach which iterates between updating the $Q$-distribution and the model parameters $\vec{\Lambda}$ (keeping the other one fixed when updating the other one). An important characteristic of the EM algorithm is that, in each iteration, the posterior distribution $Q(\vec{z})$ is selected to make the inequality in \eqref{eq:gmm-elbo} \emph{tight}, making the ELBO \emph{equal} to the log-likelihood. This is done by choosing $Q(\vec{z})$ to be the posterior distribution $P_{\vec{\Lambda}}(\vec{z}|\vec{x})$ (using the current estimates of model parameters). Importantly, this posterior can be computed in \emph{closed form}. The EM algorithm is guaranteed to converge to a local maximum of the log-likelihood. It should be noted, however, that as the likelihood function contains local maximae \cite{Jin2016-local-maxima-GMM}, global optimality is not guaranteed. The quality of the obtained GMM (in terms of log-likelihood) depends not only on the number of EM iterations, but the initial parameters.

In contrast to GMMs, the posterior distribution of VAEs \emph{cannot} be evaluated in closed form at any stage (training or scoring). For this reason, it is replaced by an \emph{approximate}, variational \cite{bishop_2006} posterior, $q_{\vec{\phi}}(\vec{z}|\vec{x})$, leading to the ELBO training objective of Eq. \eqref{eq:ELBO}. As the true posterior distribution cannot be evaluated, the EM algorithm cannot be used for VAE training \cite{kingma2013autoencoding}. The ELBO is instead optimized using gradient-based methods. Due to all these differences, it is difficult to compare VAEs and GMMs as models exactly. One of the main benefits of VAEs over GMMs is that they can handle high-dimensional inputs --- here, raw spectrograms and CQCC-grams consisting of multiple stacked frames --- allowing the use of less restrictive features.

\section{Experimental setup}

We describe our experimental setup in this section, including description of the evaluation datasets, features, model architectures and training, and performance metrics.

\subsection{Dataset}
We use the anti-spoofing dataset that has been released publicly as a result of the automatic speaker verification and spoofing countermeasures\footnote{\url{https://www.asvspoof.org/}} (ASVspoof) challenge series that started in 2013. We focus on replay attacks that are simple to mount, yet extremely challenging to detect reliably. We use the ASVspoof 2017 (version 2.0) \cite{hectorAsvspoof2.0} and ASVspoof 2019 PA \cite{asvspoof2019overview} subconditions as our evaluation data. Both datasets consist of 16 kHz audio and are complementary to each other. The former consists of \emph{real} replay recordings obtained by replaying part 1 of the \emph{RedDots} corpus \cite{redDotsDataCollection} through various devices and environments \cite{kinnunen2017reddots}. The latter dataset consists of controlled, \emph{simulated} replay attacks, detailed in \cite{asvspoof2019_evaluationplan}. Both datasets are split into three subsets: training, development and evaluation with non-overlapping speakers in each subsets. Table~\ref{dataset_statistics} summarizes the key statistics of both datasets.\\

\begin{table}[t!]
\caption{Database statistics. Spkr: speaker. Bon: bonafide/genuine, spf: spoof/replay. Each of the three subsets has non-overlapping speakers. The ASVspoof 2017 dataset has male speakers only while the ASVspoof 2019 has both male and female speakers.}
\centering
\vspace{1.5mm}
\scalebox{1.0}{
\begin{tabular}{c|ccc|ccc}
 \hline
 &\multicolumn{3}{c}{\textbf{ASVspoof 2017} \cite{tomiSummaryPaper}} &\multicolumn{3}{c} {\textbf{ASVspoof 2019 PA} \cite{asvspoof2019overview}} \\
 \hline
  
 \textbf{Subset} &\# \textbf{Spkr} & \# \textbf{Bon} & \# \textbf{Spf} & \# \textbf{Spkr} & \# \textbf{Bon} & \# \textbf{Spf} \\
 \hline
 Train &$10$ &$1507$ &$1507$  &$20$ &$5400$ &$48600$ \\
 Dev   &$8$ &$760$ &$950$     &$20$ &$5400$ &$24300$ \\
 Eval  &$24$ &$1298$ &$12008$ &$67$ &$18090$ &$116640$ \\
 \hline
 Total &$42$ &$3565$ &$14465$ &$107$ &$28890$ &$189540$ \\
 \hline

\end{tabular}}
\label{dataset_statistics}
\end{table}	

\subsection{Features and input representation}

We consider both CQCC \cite{hector_cqcc} and log-power spectrogram features. We apply a pre-processing step on the raw-audio waveforms to trim silence/noise before and after the utterance in the training, development and test sets, following recommendations in \cite{bhusanSLT2018} and \cite{bhusan2019challenge}. Following \cite{hectorAsvspoof2.0}, we extract log energy plus $19$-dimensional static coefficients augmented with deltas and double-deltas, yielding $60$-dimensional feature vectors. This is followed by cepstral mean and variance normalisation. As for the power spectrogram, we use a 512-point \emph{discrete Fourier transform} (DFT) with frame size and shift of 32 ms and 10 ms, respectively, leading to $N$ feature frames with $257$ spectral bins. We use the Librosa\footnote{\url{https://librosa.github.io/librosa/}} library to compute spectrograms. 

As our VAE models use a fixed input representation, we create a unified feature matrix by truncating or replicating the feature frames. If $N$ is less than our desired number of feature frames $T$, we copy the original $N$ frames from the beginning to obtain $T$ frames. Otherwise, if $N>T$, we retain the first $T$ frames. The point of truncating (or replicating) frames in the way described above is to ensure meaningful comparison where both models use the same audio frames as their input. This also means that the numbers reported in this paper are not directly\footnote{GMMs reported in the literature do not truncate or replicate, and this was done by us for a fair comparison with VAEs.} comparable to those reported in literature; in specific, excluding the trailing audio (mostly silence or nonspeech) after the first 1 second will increase the error rates of our baseline GMM substantially. The issue with the original, `low' error rates relates in part to database design issues, rather than bonafide/spoof discrimination \cite{bhusan2019challenge,bhusanSLT2018}. The main motivation to use the $T$ frames at the beginning is to build fixed-length utterance-level countermeasure models, which is a commonly adopted design choice for anti-spoofing systems, \emph{e.g.} \cite{galina_IS2017,zhang_JSTSP2017}.

This yields a $100\times60$-dimensional CQCC representation and a $100\times257$ power spectrogram representation for every audio file. We use the same number of frames ($T=100$) for both the GMM and VAE models.  Note that GMMs treat frames as independent observations while VAEs consider the whole matrix as a single high-dimensional data point. 

\subsection{Model architecture}\label{vae_architecture}
Our baseline GMM consists of $512$ mixture components (motivated from \cite{hectorAsvspoof2.0}) with diagonal covariance matrices. As for the VAE, our encoder and decoder model architecture is adapted from \cite{mishra2019ganbased}. For a given $T\times D$ feature matrix, where $T$=time frames and $D$=feature dimension, the encoder predicts the mean $\vec{\mu}_{z}$ and the log-variance log $\vec{\sigma}^{2}_{\vec{z}}$ that parameterize the posterior distribution $q_{\vec{\phi}}(\vec{z}|\vec{x})$, by applying a series of \emph{strided 2D convolutions} \cite{dumoulin2016guide} as detailed in Table \ref{encoder_model_architecture1_generalised}. We use a stride of 2 instead of pooling for downsampling the original input. The decoder network architecture is summarized in Table \ref{decoderr_model_architecture1_generalised}. It takes a $128$ dimensional sampled $\vec{z}$ vector as input and predicts the mean $\vec{\mu_{x}}$ and the log-variance log $\vec{\sigma}^{2}_{\vec{x}}$ that parameterize the distribution through a series of \emph{transposed convolution} \cite{dumoulin2016guide} operations. We use LeakyReLU \cite{leakyRelu} activations in all layers except the Gaussian mean and log variance layers which use linear activations. We use batch normalisation before applying non-linearity in both encoder and decoder networks.

\subsection{Model training and scoring}\label{model_training}

We train GMMs for $10$ EM iterations with random initialisation of parameters. We train bonafide and spoof GMMs separately to model the respective class distributions as discussed in Section \ref{gmm_models}. We train our VAE models (Subsection \ref{different_vae_models}) using stochastic gradient descent with Adam optimisation \cite{adam}, with an initial learning rate of $10^{-4}$ and $16$ samples as the minibatch size. We train them for $300$ epochs and stop the training if the validation loss does not improve for $10$ epochs. We apply $50\%$ dropout to the inputs of fully connected layers in our auxiliary classifier. We do not apply dropout in the encoder and decoder network.

\subsection{Performance measures}
We assess the bonafide-spoof detection performance in terms of the \emph{equal error rate} (EER) of each countermeasure. EER was the primary evaluation metric of the ASVspoof 2017 challenge, and a secondary metric of the ASVspoof 2019 challenge. EER is the error rate at an operating point where the false acceptance (\emph{false alarm}) and false rejection (\emph{miss}) rates are equal. A reference value of 50\% indicates the chance level.

In addition to EER, which evaluates countermeasure performance in isolation from ASV, we report the \emph{tandem detection cost function} (t-DCF) \cite{tomi_tDCF} which evaluates countermeasure and ASV performance jointly under a Bayesian decision risk approach. We use the same t-DCF cost and prior parameters as used in the ASVspoof2019 evaluation \cite{asvspoof2019_evaluationplan}, with the x-vector probabilistic linear discriminant analysis (PLDA) scores provided by the organizers of the same challenge. The ASV system is set to its EER operating point while the (normalized) t-DCF is reported by setting the countermeasure to its minimum-cost operating point. We report both metrics using the official scripts released by the organizers. A reference value 1.00 of (normalized) t-DCF indicates an uninformative countermeasure.

\begin{table}[t!]
	\caption {Encoder model architecture. Conv denotes a convolutional operation. T: number of time frames. F: number of feature dimensions. The scalar f is the size of the flattened vector from the last Conv layer output, and represents the number of input units to $\vec{\mu_{\vec{z}}}$ and log $\vec{\vec{\sigma}}^{2}_{\vec{z}}$ fully connected layers. M=$16$ for spectrogram inputs and $32$ for CQCCs. Conv 5 layer is not applicable for CQCCs.}
	\centering
	\vspace{1.5mm}
	\scalebox{0.96}{
	\begin{tabular}{cccccc}
		\hline
		\textbf{Layer}  &\textbf{Input} &\textbf{Filter} &\textbf{Stride}  &\# \textbf{Filters/} &\textbf{Output} \\
		\textbf{}  &\textbf{shape} &\textbf{size} &\textbf{size}  &\textbf{neurons} &\textbf{shape} \\
		\hline
		Conv 1 &T$\times$F$\times$1 &5$\times$257 &2$\times$2 &M &T/2$\times$F/2$\times$M \\
		Conv 2 &T/2$\times$F/2$\times$M &5$\times$129 &2$\times$2 &2M &T/4$\times$F/4$\times$2M \\
		Conv 3 &T/4$\times$F/4$\times$2M &5$\times$65 &2$\times$2 &4M &T/8$\times$F/8$\times$4M \\
		Conv 4 &T/8$\times$F/8$\times$4M &5$\times$33 &2$\times$2 &8M &T/16$\times$F/16$\times$8M \\
		Conv 5 &T/16$\times$F/16$\times$8M &5$\times$17 &2$\times$2 &16M &T/32$\times$F/32$\times$16M \\
		
		$\vec{\mu_{z}}$ &f &-&- &$128$ &$128$ \\				
		log $\vec{\sigma}^{2}_{\vec{z}}$ &f &-&- &$128$ &$128$ \\
		\hline		
	\end{tabular}}
		\label{encoder_model_architecture1_generalised}
\end{table}

\begin{table}[t!]
\caption{Decoder model architecture. ConvT denotes a transposed convolutional operation. * denotes zero padding operation applied to match the input shape. The Gaussian layers $\vec{\mu_{\vec{x}}}$ and log $\vec{\vec{\sigma}}^{2}_{\vec{x}}$ use Conv operation. The initial values of H and W depend on the number of neurons (\#neurons) in the FC layer which is $12288$ for spectrograms and $2304$ for CQCCs.}
	\centering
	\vspace{1.5mm}
	\scalebox{0.96}{
	\begin{tabular}{cccccc}
	\hline
	\textbf{Layer}  &\textbf{Input} &\textbf{Filter} &\textbf{Stride}  &\# \textbf{Filters} &\textbf{Output} \\
	\textbf{}  &\textbf{shape} &\textbf{size} &\textbf{size}  &\textbf{neurons} &\textbf{shape} \\
	\hline
	FC &$128$ & - &- & \#neurons & \#neurons \\
	ConvT &H$\times$W$\times$128 &5$\times$10 &2$\times$2 &64 &2H$\times$2W$\times$64 \\
	ConvT &2H$\times$2W$\times$64 &5$\times$20 &2$\times$2 &32 &4H$\times$4W$\times$32 \\
	ConvT* &5H$\times$4W$\times$32 &5$\times$20 &2$\times$2 &16 &10H$\times$8W$\times$16 \\
	ConvT* &10H$\times$8W$\times$16 &5$\times$20 &2$\times$2 &8 &20H$\times$16W$\times$8 \\
	
	$\vec{\mu_{x}}$* &100$\times$F$\times$8 &5$\times$5 &1$\times$1 &1 &100$\times$F$\times$1 \\
	
	log $\vec{\sigma}^{2}_{\vec{x}}*$ &100$\times$F$\times$8 &5$\times$5 &1$\times$1 &1 &100$\times$F$\times$1 \\ 
	\hline	
	
\end{tabular}}
\label{decoderr_model_architecture1_generalised}
\end{table}

\subsection{Experiments}

We perform several experiments using different VAE setups using CQCCs and log-power spectrogram inputs as described in Subsection \ref{different_vae_models}. We also train baseline GMMs for comparing VAE performance using the same input CQCC features. While training VAEs with an auxiliary classifier on the $\vec{\mu}_{\vec{z}}$ input, we use $32$ neuron units on the FC layer. We do not use the entire training and development audio files for training and model validation on the ASVspoof 2019 dataset, but adopt custom training and development protocols used in \cite{bhusan2019challenge} that showed good generalisation on the ASVspoof 2019 test dataset during the recent ASVspoof 2019 evaluations. Note, however, that all the evaluation portion results are reported on the standard ASVspoof protocols. In the next section we describe our experimental results.

\section{Results and discussion}

\subsection{Impact of latent space dimensionality}

\begin{table}[t!]
	\caption{EER vs dimension of latent space. Showing the effect of latent dimension on the performance metric for the C-VAE model when trained using CQCCs and spectrograms. Shown results are on the evaluation set. Shown results are on the ASVspoof 2017 dataset.}
	\centering
	\vspace{1.5mm}
	\scalebox{1.0}{
		\begin{tabular}{ccc|ccc}
			\hline
			&\multicolumn{2}{c}{Spectrogram} &\multicolumn{2}{|c}{CQCC} \\
			\cline{2-5}
			Latent Dimension & EER & t-DCF & EER & t-DCF \\
			\hline
			$8$ &$31.20$ &$0.8642$  &$33.35$ &$0.8584$ \\
			$16$ & $26.88$ &$0.7551$  &$33.74$ &$0.8542$ \\
			$32$ &$36.65$ &$0.9383$ & $30.81$ &$0.7909$ \\
			$64$ &$29.73$ &$0.7650$ &$29.52$ &$0.7325$ \\
			$128$ &$29.43$ &$0.7303$ &$29.27$ &$0.7222$ \\
			$256$ &$29.80$ &$0.7609$ &$28.87$ &$0.6962$ \\
			$512$ &$25.73$ &$0.6662$ &$28.42$ &$0.7033$ \\
			\hline
		\end{tabular}}
		\label{eer_vs_z_dimension_spect_input}
\end{table}

We first address the impact of latent space dimensionality on the ASVspoof 2017 corpus. To keep computation time manageable, we focus only on the C-VAE variant. The results, for both CQCC and spectrogram features, are summarized in Table \ref{eer_vs_z_dimension_spect_input}. We observe an overall decreasing trend in EER with increased latent space dimensionality, as expected. All the error rates are relatively high, which indicates general difficulty of our detection task. In the remainder of this study, we fix the latent space dimensionality to $d=128$ as a suitable trade-off in EER and computation.

\subsection{Comparing the performance of different VAE setups with GMM} 

Our next experiment addresses the relative performance of different VAE variants and their relation to our GMM baseline. As GMMs cannot be used with high-dimensional spectrogram inputs, the results are shown only for the CQCC features. This experiment serves to answer the question on which VAE variants are the most promising, and whether VAEs could be used to replace the standard GMM as a back-end classifier. The results for both the ASVspoof 2017 and 2019 (PA) datasets are summarized in Table~\ref{main_results_table}. \\

\noindent
\textbf{Baseline GMM.} On the ASVspoof 2017 dataset, the GMM displays EERs of $19.07$\% and $22.6$\% on the development and evaluation sets, respectively. Note that our baseline is completely different from the CQCC-GMM results of \cite{hectorAsvspoof2.0} for two reasons. First, we use a unified time representation of the first $100$ frames obtained either by truncating or copying time frames, for reasons explained earlier. Second, we remove the leading and trailing nonspeech/silence from every utterance, to mitigate a dataset-related bias identified in \cite{bhusanSLT2018}: the goal of our modified setup is to ensure that our models focus on actual factors, rather than database artefacts. 

\begin{table}[t!]
	\caption{Performance of GMM (baseline) and different VAE models using \textbf{CQCCs} as input feature. AC-VAE$_1$: augmenting classifier on top of the latent space. AC-VAE$_2$: augmenting classifier at the output of the decoder. Highlighted in bold indicates the best performance among VAE variants. }
	\centering
	\vspace{1.5mm}
	\scalebox{0.95}{
		\begin{tabular}{ccccc|cccc}
			\hline
			&\multicolumn{4}{c}{ASVspoof 2017} &\multicolumn{4}{c}{ASVspoof 2019 PA} \\
			\cline{2-9}
			&\multicolumn{2}{c}{Dev} &\multicolumn{2}{c}{Eval} &\multicolumn{2}{c}{Dev} &\multicolumn{2}{c}{Eval} \\
			\cline{2-9}
			Model &EER &t-DCF &EER &t-DCF &EER &t-DCF &EER &t-DCF \\
			\hline
			GMM &$19.07$ &$0.4365$ &$22.6$ &$0.6211$ &$43.77$ &$0.9973$ &$45.48$ &$0.9988$ \\
			VAE &$29.2$ &$0.7532$ &$32.37$ &$0.8079$ &$45.24$ &$0.9855$ &$45.53$ &$0.9978$ \\
			C-VAE &$18.1$ &$0.4635$ &$\textbf{28.1}$ &$\textbf{0.7020}$ &$\textbf{34.06}$ &$\textbf{0.8129}$ &$36.66$ &$0.9104$ \\
			AC-VAE$_1$ &$21.8$ &$0.4914$ &$29.3$ &$0.7365$ &$34.73$ &$0.8516$ &$36.42$ &$0.9036$ \\
			AC-VAE$_2$ &$\textbf{17.78}$ &$\textbf{0.4469}$ &$29.73$ &$0.7368$ &$34.87$ &$0.8430$ &$\textbf{36.42}$ &$\textbf{0.8963}$ \\
			\hline
		\end{tabular}}
		\label{main_results_table}
\end{table}

On the ASVspoof 2019 PA dataset, the performance of the GMM baseline\footnote{For sanity check, we trained a GMM without removing silence (and using all frames per utterance) and obtained a performance similar to the official GMM baseline of the ASVspoof 2019 challenge. On the development set, our GMM now shows an EER of $10.06$\% and t-DCF of $0.1971$ which is slightly worse than official baseline (EER = $9.87$ and t-DCF =$0.1953$).} is nearly random as indicated by both metrics. The difficulty of the task and our modified setup to suppress database artifacts both contribute to high error rates. The results are consistent with our earlier findings in \cite{bhusan2019challenge}. The two separate GMMs may have learnt similar data distributions. Note that the similarly-trained naive VAE displays similar near-random performance. \\  

\noindent
\textbf{VAE variants.} Let us first focus on the ASVspoof 2017 results. Our first, naive VAE approach falls systematically behind our baseline GMM. Even if both the bonafide and spoof VAE models display decent reconstructions, they lack the ability to retain discriminative information in the latent space when trained in isolation from each other. Further, remembering that VAE training optimizes only a lower bound of the log-likelihood function, it might happen that the detection score formed as a difference of these `inexact' log-likelihoods either under- or overshoots the true log-likelihood ratio --- but there is no way of knowing which way it is.

The C-VAE model, however, shows encouraging results compared with all the other VAE variants considered. This suggests that conditioning both the encoder and decoder with class labels during VAE training is helpful. Supposedly a shared, conditional C-VAE model yields `more compatible' bonafide and spoof scores when we form the detection score. The C-VAE model shows comparable detection performance to the GMM baseline on the development set, though it performs poorly on the evaluation set. 

\begin{table}[t!]
	\caption{Comparing VAE and C-VAE performance on the ASVspoof 2017 dataset using the log power spectrograms as input features. }
	\centering
	\vspace{1.5mm}
	\scalebox{1.0}{
		\begin{tabular}{ccccc}
			\hline
			&\multicolumn{2}{c}{Dev} &\multicolumn{2}{c}{Eval} \\ 
			\cline{2-5}
			Model &EER &t-DCF &EER &t-DCF \\ 
			\hline
			VAE  &$32.12$ &$0.8037$ &$36.92$ &$0.9426$ \\
			C-VAE &$22.81$ &$0.5219$ &$29.52$ &$0.7302$ \\
			\hline
		\end{tabular}}
		\label{results_comparing_vae_cvae_on_spectrograms}
\end{table}

The VAE variants with an auxiliary classifier outperform the naive VAE but are behind C-VAE: both AC-VAE$_{1}$ and AC-VAE$_2$ display slightly degraded performance over C-VAE on the evaluation set. While AC-VAE$_1$ and AC-VAE$_2$ show comparable performance on the evaluation set, on the development set AC-VAE$_2$ outperforms all other VAE variants in both metrics. This suggests overfitting on the development set: adding an auxiliary classifier increases the model complexity as the number of free parameters to be learned increases substantially. Apart from having to learn optimal model parameters from a small training dataset, another challenge is to find an optimal value for the control parameters $\alpha$ and $\beta$ in \eqref{aux_classifier_loss}. 

On the ASVspoof 2019\footnote{We would like to stress that we do not use the original training and development protocols for model training and validation. Instead, we use custom, but publicly released protocols available at \url{https://github.com/BhusanChettri/ASVspoof2019} from our prior work \cite{bhusan2019challenge} that helped to improve generalisation during the ASVspoof 2019 challenge. However, during testing, we report test results on the standard development and evaluation protocols.} dataset our C-VAE model now outperforms the naive VAE and the GMM baseline. By conditioning the encoder and decoder networks with class labels, we observe an absolute improvement of about $10$\% over the naive VAE on both the development and the evaluation sets. Unlike in the ASVspoof 2017 dataset, the auxiliary classifier VAE now offers some improvement on the evaluation set. This might be due to much larger number of training examples available in the ASVspoof 2019 dataset  ($54000$ utterances) in comparison to the ASVspoof 2017 training set ($3014$ utterances).

It should be further noted that while training models on ASVspoof 2019 dataset, we used the hyper-parameters (learning rate, mini-batch size, control parameters including the network architecture) that were optimised on the ASVspoof 2017 dataset. This was done to study how well the architecture and hyper-parameters generalize from one replay dataset (ASVspoof 2017) to another one (ASVspoof 2019).

The results in Table \ref{main_results_table} with the CQCC features indicate that the C-VAE is the most promising variant for further experiments. While adding the auxiliary classifier improved performance in a few cases, the improvements are modest relative to the added complexity. Therefore, in the remainder of this paper, we focus on the C-VAE unless otherwise stated. Also, we focus testing our ideas on the ASVspoof 2017 replay dataset for computational reasons. Next, to confirm the observed performance improvement of C-VAE over naive VAE, we further train both models using raw log power-spectrogram features. The results in Table \ref{results_comparing_vae_cvae_on_spectrograms} confirm the anticipated result in terms of both metrics.

\begin{figure}[t!] %[t]
	\centering  
	\includegraphics[width=\linewidth]{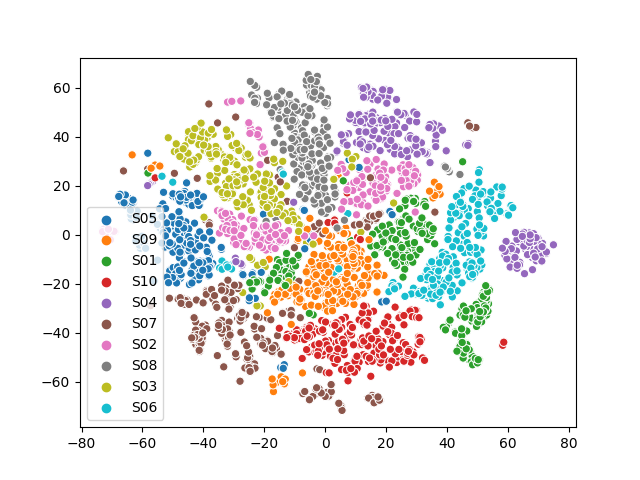}
	%\vspace{-2.7cm}
	\caption{Visualisation of the latent space for $10$ different sentences in the ASVspoof 2017 training set by C-VAE with genuine-class conditioning. S01: `My voice is my password'. S02: `OK Google'. S03: `Only lawyers love millionaires'. S04: `Artificial intelligence is for real'. S05: `Birthday parties have cupcakes and ice cream'. S06: `Actions speak louder than words'. S07: `There is no such thing as a free lunch'. S08: `A watched pot never boils'. S09: `Jealousy has twenty-twenty vision'. S10: `Necessity is the mother of invention'. }
	\label{tsne_visuals_2017_sentences}
\end{figure}

\subsection{Conditioning VAEs beyond class labels}

The results so far confirm that the C-VAE outperforms the naive VAE by a wide margin. We now focus on multi-class conditioning using C-VAEs. To this end, our possible conditioning variables could include speaker and sentence identifiers. However, speakers are different across the training and test sets in both  ASVspoof 2017 and ASVspoof 2019, preventing the use %Therefore, multi-class conditioning on 
of speaker conditioning. Further, the phrase identities of the ASVspoof 2019 PA dataset are not publicly available. For these reasons we restrict our focus on the $10$ common passphrases in the ASVspoof 2017 dataset shared across training, development and evaluation data. The contents of each phrase (S01 through S10) are provided in the caption of Fig.~\ref{tsne_visuals_2017_sentences}. The number of bonafide and spoof utterances for these passphrases in the training and development sets are equally balanced. We therefore use a $20$-dimensional one-hot vector to represent multi-class condition. The first $10$ labels correspond to bonafide sentences S01 through S10 and the remaining $10$ to spoofed utterances. Everything else about training and scoring the C-VAE model remains the same as above, except for the use of the $20$-dimensional (rather than $2$-dimensional) one-hot vector. 

We first visualise how the latent space is distributed across the $10$ different phrases of the ASVspoof 2017 training set. Fig.~\ref{tsne_visuals_2017_sentences} shows the t-SNE \cite{tsne_Maaten2008VisualizingDU} plots for $10$ different utterances in the ASVspoof 2017 dataset. The clear distinction between different phrases suggests that the latent space preserves the structure and identity of different sentences of the dataset. This suggests that choosing the sentence identity for conditioning the VAE might be beneficial towards improving performance; such model is expected to learn \emph{phrase-specific} bonafide-vs-spoof discriminatory cues.

Table \ref{comparing_cvae_two_class_20_class_conditioning_new} summarises the results. The C-VAE trained on spectrogram features with multi-class conditioning shows a substantial improvement over two-class conditioning. This suggests that the network now benefits exploiting relevant information present across different passphrases, which may be difficult from binary class conditioning. For the CQCC features, however, we have the opposite finding: while EER is slightly decreased on the evaluation set with multi-class conditioning, overall it shows degraded performance. One possible interpretation is that CQCCs are a compact feature representation optimized specifically for anti-spoofing. CQCCs may lack phrase-specific information relevant for anti-spoofing which is retained by the richer and higher-dimensional raw spectrogram. To sum up, the C-VAE trained on raw spectrograms with multi-class conditioning offers substantial improvement over two-class conditioning in comparison to CQCC input features.

\begin{table}[t!]
	\caption{Comparing the performance of the C-VAE model trained using two-class and multi-class (10 bonafide and 10 spoof phrases) conditioning. Shown results are on the ASVspoof 2017 v2.0 dataset using CQCC and spectrogram inputs. For two-class conditioning we simply use the class labels yielding 2 dimensional one-hot vector. For multi-class conditioning we use the $10$ different passphrases of the ASVspoof 2017 v2.0 dataset. This results in a $20$ dimensional one-hot vector ($10$ for bonafide and $10$ for spooof). The results for two-class conditioning (the first row) are included from Table~\ref{main_results_table} and \ref{results_comparing_vae_cvae_on_spectrograms} for better readability.}
	\centering
	\vspace{1.5mm}
	\scalebox{0.95}{
		\begin{tabular}{ccccc|cccc}
			\hline
			&\multicolumn{4}{c}{CQCC} &\multicolumn{4}{c}{Spectrogram} \\
			\cline{2-9}
			&\multicolumn{2}{c}{Dev} &\multicolumn{2}{c}{Eval} &\multicolumn{2}{c}{Dev} &\multicolumn{2}{c}{Eval} \\
			\cline{2-9}
			Conditioning &EER &t-DCF &EER &t-DCF &EER &t-DCF &EER &t-DCF \\
			\hline
			
			Two-class  &$18.1$ &$0.4635$ &$28.1$ &$0.7020$ &$22.81$ &$0.5219$ &$29.52$ &$0.7302$ \\
			
			Multi-class &$19.77$ &$0.4961$ &$27.88$ &$0.7390$ &$19.65$ &$0.4324$ &$25.48$ &$0.6631$ \\
			
			\hline
			
		\end{tabular}}
		\label{comparing_cvae_two_class_20_class_conditioning_new}
\end{table}

\begin{figure}[htb]
	\begin{minipage}[b]{.48\linewidth}
		\centering
		\centerline{\includegraphics[width=6.8cm]{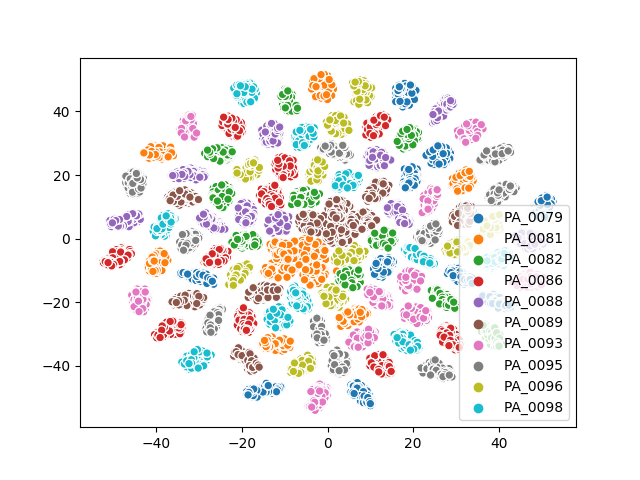}}
	
	\end{minipage}
	\hfill
	\begin{minipage}[b]{0.48\linewidth}
		\centering
		\centerline{\includegraphics[width=6.8cm]{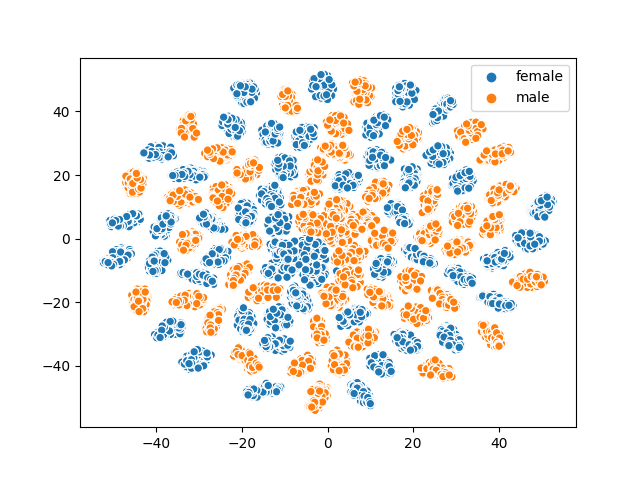}}
		
	\end{minipage}
	
	\begin{minipage}[b]{.48\linewidth}
		\centering
		\centerline{\includegraphics[width=6.8cm]{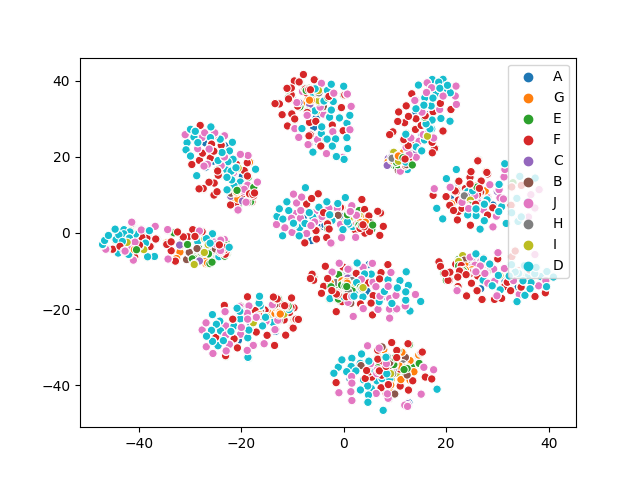}}
		
	\end{minipage}
	\hfill
	\begin{minipage}[b]{0.48\linewidth}
		\centering
		\centerline{\includegraphics[width=6.8cm]{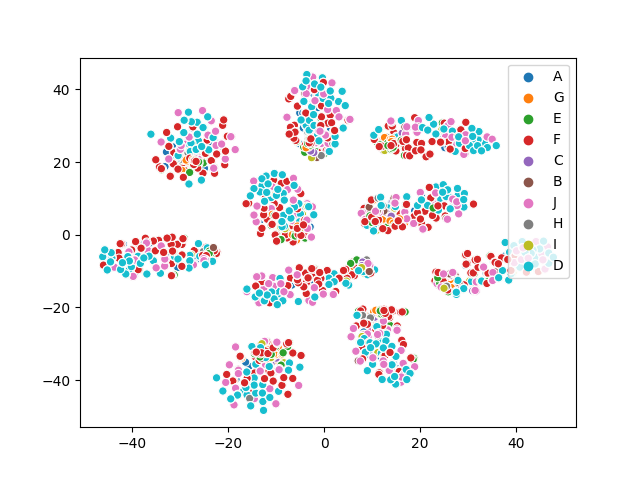}}
		
	\end{minipage}
	\caption{t-SNE visualisations of latent space for $5$ male and $5$ female speaker utterances drawn randomly from the ASVspoof 2019 PA training set. (a) \textbf{Top left}: represents $10$ different speaker identities. (b) \textbf{Top right}: male and female clusters (c) \textbf{Bottom left}: distribution of  bonafide and attack conditions for a male speaker PA$\_0082$. (d) \textbf{Bottom right}: same as in (c) but for a female speaker PA$\_0079$. A-I indicate bonafide and $9$ different attack conditions whose original labels are as follows. A: bonafide, B: `BB', C: `BA', D: `CC', E: `AB', F: `AC', G: `AA', H: `CA', I: `CB', J: `BC'. See \cite{asvspoof2019overview} for details of these labels.}
	\label{tsne_plots_2019}
\end{figure}

\subsection{Qualitative results}
A relevant question is whether or not the latent space features $\vec{z}$ have some clear meaning in terms of human or spoofed speech parameters, or any other relevant information that helps us derive some understanding about the underlying data. To this end, we analyse the latent space through 2D visualisations using the t-SNE algorithm. We aim to understand how the latent space is distributed across different speakers and between genders. We do this on the ASVspoof 2019 dataset, as the 2017 dataset only has male speakers. Fig.~\ref{tsne_plots_2019} shows t-SNE plots for $5$ male and $5$ female speakers on the ASVspoof 2019 PA training set chosen randomly.

Subfigures in the first row of Fig.~\ref{tsne_plots_2019} suggest that the latent space has learned quite well to capture speaker and gender specific information. We further analyse bonafide and different attack conditions per gender, taking PA\_0082 and PA\_0079 --- one male and female speaker randomly from the pool of $10$ speakers we considered. Fig.~\ref{tsne_plots_2019}, second row illustrates this. We use letters A-I to indicate bonafide and $9$ different attack conditions whose original labels are as follows. A: bonafide, B: `BB', C: `BA', D: `CC', E: `AB', F: `AC', G: `AA', H: `CA', I: `CB', J: `BC'. See \cite{asvspoof2019overview} for details of these labels. From Fig.~\ref{tsne_plots_2019}, we observe overlapping attacks within a cluster, and spread of these attacks across different clusters. The bonafide audio examples, denoted by letter A are heavily overlapped by various spoofed examples. This gives an intuition that the latent space is unable to preserve much discriminative information due to the nature of the task, and rather, it might be focusing on generic speech attributes such as acoustic content, speaker speaking style to name a few, to be able to generate a reasonable reconstruction --- as depicted in Fig~\ref{vae_figures_spectrograms_cvae}.

\subsection{VAE as a feature extractor}

\begin{figure}
	\subfloat[Audio: D\_1000022 with bonafide-class conditioning]{\includegraphics[width=2.5in]{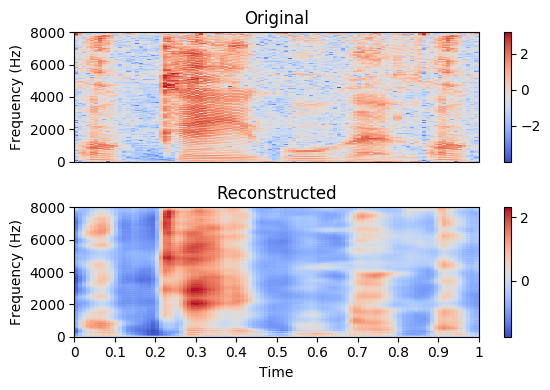}} 
	\subfloat[Audio: D\_1000022 with spoof-class conditioning]{\includegraphics[width=2.5in]{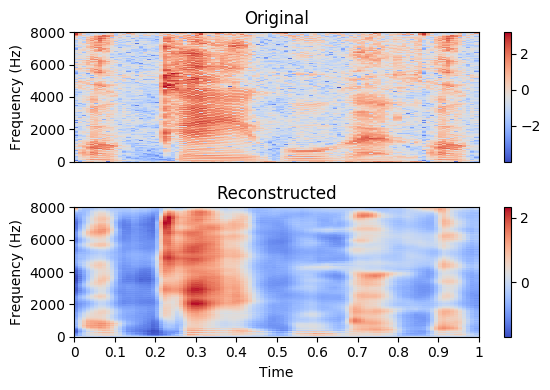}}\\
	\subfloat[Audio: D\_1001049 with bonafide-class conditioning]{\includegraphics[width=2.5in]{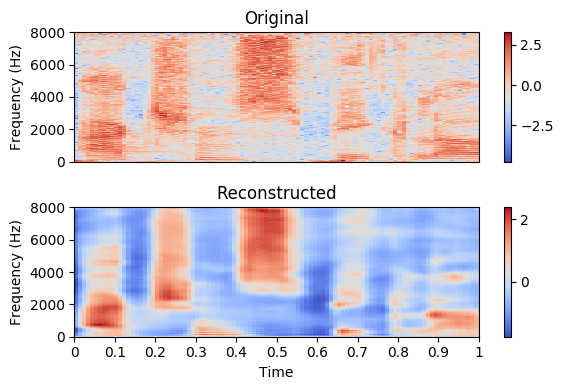}}
	\subfloat[Audio: D\_1001049 with spoof-class conditioning]{\includegraphics[width=2.5in]{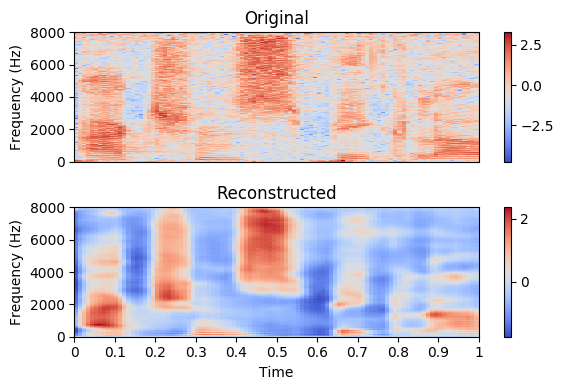}} \\
	\caption{Visualisation of the reconstructed spectrograms by the  C-VAE. Shown are the reconstruction of bonafide (D\_1000022) and spoof (D\_1001049) audio examples using bonafide and spoof class conditioning respectively. The audio examples are taken from the ASVspoof 2017 development set.}
	\label{vae_figures_spectrograms_cvae}
\end{figure}

The results shown in Fig. \ref{vae_figures_spectrograms_cvae} indicate that our VAEs have learnt to reconstruct spectrograms using prominent acoustic cues and, further, the latent codes visualized in Fig. \ref{tsne_visuals_2017_sentences} indicate strong content dependency. The latent space in a VAE may therefore focus on retaining information such as broad spectral structure and formants that help in increasing the data likelihood leading to good reconstruction. But in spoofing attack detection (especially the case of high-quality replay attacks) we are also interested in \emph{detail} --- the part \emph{not} modeled by a VAE. This lead the authors to consider an alternative use case of VAE as a \emph{feature extractor}. 

\begin{figure}[t!]
	\centering  
	\includegraphics[width=\linewidth]{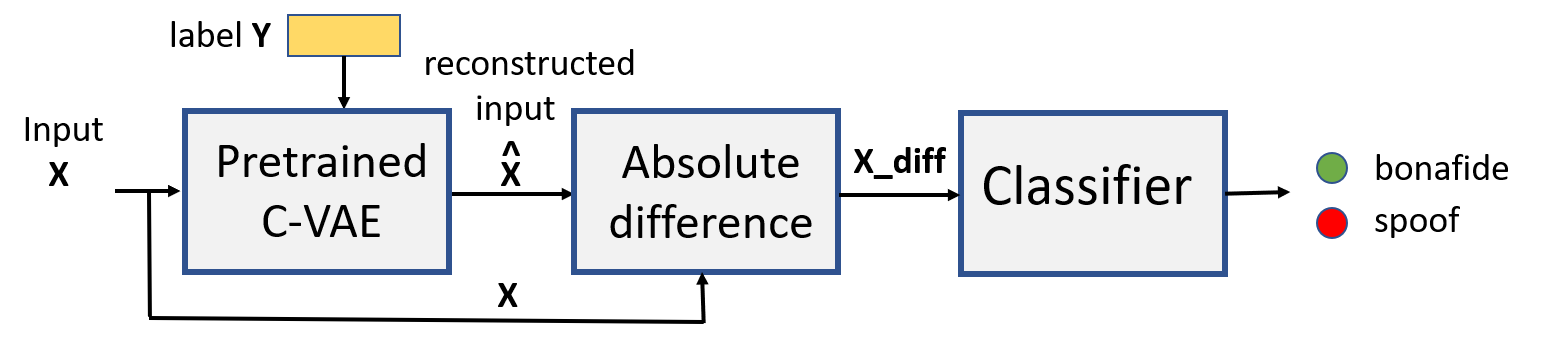}
	\caption{VAE as a feature extractor. A pretrained C-VAE model first produces a reconstructed input. The difference of the original and reconstructed input is used as a new feature representation. This feature representation is obtained for the entire dataset and a new classifier (CNN in this case) is trained on the training set features. The development set is used for model validation.}
	\label{fig:VAE-as-feature-extractor}
\end{figure}

The idea is illustrated in Fig. \ref{fig:VAE-as-feature-extractor}. We use our pre-trained C-VAE model (with bonafide-class conditioning) to obtain a new feature representation that we dub as \textbf{VAE residual}, defined as the absolute difference of the input spectrogram and the reconstructed spectrogram by the C-VAE model. We extract the VAE residual features from all training utterances and train a new classifier back-end (here, a CNN) using these features as input. We adopt the CNN architecture and training from \cite{bhusanSLT2018}. During testing, we use the CNN output activation (sigmoid activation) as our spoof detection score. Though another recent study also used VAEs for feature extraction \cite{yexin2019challenge}, our approach is different; the authors of \cite{yexin2019challenge} used the latent variable from a pretrained VAE model, while we use the residual of the original and reconstructed inputs.

\begin{table}[t!]
	\caption{C-VAE performance comparison under different settings. Results shown in the first row are taken from Table \ref{results_comparing_vae_cvae_on_spectrograms} second row for comparison --- C-VAE trained on spectrograms. The second row shows results when the same C-VAE model (with bonafide class conditioning) is used as a feature extractor. The third row shows the results when the same CNN classifier is trained on spectrogram input so as to see how it compares with the one trained on VAE residuals. Results shown are on ASVspoof 2017 dataset.}
	\centering
	\vspace{1.5mm}
	\scalebox{1.0}{
		\begin{tabular}{cccccc}
			\hline
    		&&\multicolumn{2}{c}{Dev} &\multicolumn{2}{c}{Eval}  \\
			\cline{3-6}
			Features &Model &EER &t-DCF &EER &t-DCF  \\
			\hline
			Spectrogram &C-VAE &$22.81$ &$0.5219$ &$29.52$ &$0.7302$  \\
			
			VAE residual &CNN &$13.16$ &$0.3438$ &$17.32$ &$0.4293$  \\
			
			Spectrogram &CNN &$10.82$ &$0.2877$ &$16.03$ &$0.4461$ \\
			\hline
		\end{tabular}}
		\label{spectral_difference_as_a_feature_for_cnn}
\end{table}

Table \ref{spectral_difference_as_a_feature_for_cnn} summarizes the results. Numbers in the second row correspond to our proposed approach of using VAE residual features and training a separate classifier. We also include C-VAE results from initial approach (C-VAE as a back-end) from the third row of Table \ref{main_results_table} for comparison. For contrastive purposes, we train another CNN classifier (using the same architecture) using the original spectrogram directly. Using VAE residuals and training a separate classifier outperforms the back-end approach on both metrics and on both the development and evaluation sets. The residual approach, however, remains behind the CNN trained directly on the original spectrogram, on the development set. On the evaluation set, it achieves the lowest t-DCF and displays a comparable EER. The small performance gap (in relative terms) between the development and evaluation sets for the VAE residual approach suggests good generalisation.

Although the proposed VAE residual approach did not outperform the raw-spectrogram CNN, the results obtained are encouraging and show potential for further investigation. In fact, given the similar performance of the original and VAE residual spectrogram features, we interpret the results to mean that \emph{most of} the relevant information for discriminating bonafide and replay utterances (on this data) lies in the residual or `noise' part of the spectrogram. It is noteworthy that heuristic ideas inspired directly by simple visualizations such as Figs. \ref{tsne_visuals_2017_sentences} and \ref{vae_figures_spectrograms_cvae} lead to boosted performance. Finally, recalling our initial motivations, VAE leads to a generative model (unlike CNN) that allows data sampling and obtaining uncertainty of the latent space representation. These favorable properties of VAEs suggest further studies towards more versatile spoofing countermeasure solutions where the semantics, sanity and stability of the learned feature representation can be easily explored.

\section{Conclusions and future work}

Inspired by the successful use of GMMs, a classical generative model, as a backend in spoofing detection for ASV, we performed a feasibility study of using an alternative generative model -- \textit{deep generative} VAEs -- as a backend classifier. Our first study using two separate VAEs suggests that it is difficult to capture discriminative information when the models are trained using only one-class data. Both the bonafide and spoof VAEs seem to focus on retaining information relevant for data reconstruction while giving less attention on class-discriminative information. As a result, the latent space in both bonafide and spoof VAEs appears to capture common prominent characteristics of bonafide and spoofed speech, making the detection task difficult. Nonetheless, our qualitative results indicate that both our bonafide and spoof VAEs yield reasonable reconstruction of the input data.

Our second approach of training a single conditional VAE (C-VAE) by conditioning the encoder and decoder networks by class-label vectors shows far more encouraging results. The performance of our C-VAE models on both the ASVspoof 2017 and ASVspoof 2019 datasets show remarkable improvement in comparison to the naive VAE approach. Using an auxiliary classifier (AC) did not help much. We did not observe substantial improvement in detection performance on the ASVspoof 2017 dataset, though we observed some performance gain on the ASVspoof 2019 dataset, suggesting the importance  of training set size for improved generalisation. 

Despite different dataset sizes in the ASVspoof 2017 and ASVspoof 2019 datasets, we find that the model hyper-parameters tuned on the ASVspoof 2017 dataset worked quite well when applied on the 2019 dataset, showing consistency of our findings with C-VAE models. However, optimisation of network architecture and model hyper-parameters has not been fully explored in the present study, leaving scope for further improvements.

To sum up, based on both the observed detection performance and architecture complexity considerations, from the three VAE back-end variants considered (Fig. \ref{vae_figures}), the authors recommend potential future work to focus on conditional VAE (C-VAE). In fact, we obtained promising results by further conditioning C-VAE using pass-phrase label. This  warrants future studies with other conditioning variables such as speaker identity, gender, channel. Frame-level phone labels obtained through forced alignment procedure might also be interesting with alternative frame-by-frame VAE architectures.

Our primary focus has been to study the feasibility of VAE as a back-end classifier, but we also included a preliminary study on an alternative use case for spectrogram residual feature extraction. The front-end approach demonstrated substantial improvement over the VAE back-end use case, which warrants further studies. In future work, we also plan to focus on investigating alternative architectures for encoder and decoder networks involving recurrent layers. Experiments for the detection of text-to-speech and voice-conversion attacks would be interesting as well.

\section*{Acknowledgements}
This work was supported in part by the Academy of Finland (Proj. No. 309629 --- entitled “NOTCH: NOn-cooperaTive speaker CHaracterization”). EB is supported by RAEng Research Fellowship RF/128 and a Turing Fellowship. We gratefully acknowledge the support of NVIDIA Corporation with the donation of the Titan V GPU used for this research.

%\section*{References}

\bibliography{bibliography}

\begin{thebibliography}{10}
\expandafter\ifx\csname url\endcsname\relax
  \def\url#1{\texttt{#1}}\fi
\expandafter\ifx\csname urlprefix\endcsname\relax\def\urlprefix{URL }\fi
\expandafter\ifx\csname href\endcsname\relax
  \def\href#1#2{#2} \def\path#1{#1}\fi

\bibitem{reynolds_SC1995}
D.~A. Reynolds, Speaker identification and verification using {G}aussian
  mixture speaker models, Speech communication 17~(1) (1995) 91--108.

\bibitem{sahid_PAD_book}
M.~Sahidullah, H.~Delgado, M.~Todisco, T.~Kinnunen, N.~Evans, J.~Yamagishi,
  K.-A. Lee, Introduction to {V}oice {P}resentation {A}ttack {D}etection and
  {R}ecent {A}dvances (2019).
\newblock \href {http://arxiv.org/abs/1901.01085} {\path{arXiv:1901.01085}}.

\bibitem{iso_spoofing_standards}
{ISO/IEC 30107-1:2016},
  \href{https://www.iso.org/obp/ui/#iso:std:iso-iec:30107:-1:ed-1:v1:en.}{Information
  technology - {B}iometric presentation attack detection - part 1: Framework}
  (2016).
\newline\urlprefix\url{https://www.iso.org/obp/ui/#iso:std:iso-iec:30107:-1:ed-1:v1:en.}

\bibitem{Masuko99onthe}
T.~Masuko, T.~Hitotsumatsu, K.~Tokuda, T.~Kobayashi, On {T}he {S}ecurity of
  {H}mm-{B}ased {S}peaker {V}erification {S}ystems {A}gainst {I}mposture
  {U}sing {S}ynthetic {S}peech, in: In Proceedings of the European Conference
  on Speech Communication and Technology, 1999, pp. 1223--1226.

\bibitem{PellomH99}
B.~L. Pellom, J.~H.~L. Hansen, An experimental study of speaker verification
  sensitivity to computer voice-altered imposters, in: Proc. {ICASSP}, March
  15-19, 1999, pp. 837--840.

\bibitem{wu_APSIPA2014}
Z.~Wu, S.~Gao, E.~S. Cling, H.~Li, A study on replay attack and anti-spoofing
  for text-dependent speaker verification, in: Asia-Pacific Signal and
  Information Processing Association, 2014 Annual Summit and Conference
  (APSIPA), IEEE, 2014, pp. 1--5.

\bibitem{lau_2004}
L.~Y. W., W.~M., T.~D., Vulnerability of speaker verification to voice
  mimicking, in: Proc. of International symposium on Intelligent Multimedia,
  Video \& Speech Processing, Hongkong, 2004.

\bibitem{khoury_ivec_antispoofing}
E.~Khoury, T.~Kinnunen, A.~Sizov, Z.~Wu, S.~Marcel, {I}ntroducing i-vectors for
  joint anti-spoofing and speaker verification, in: Proc. INTERSPEECH, 2014,
  pp. 61--65.

\bibitem{novoseloy_IS2015}
S.~Novoselov, A.~Kozlov, G.~Lavrentyeva, K.~Simonchik, V.~Shchemelinin, {STC}
  anti-spoofing systems for the {ASV}spoof 2015 challenge, in: Proc. ICASSP,
  2016, pp. 5475--5479.

\bibitem{jennifer2019challenge}
J.~Williams, J.~Rownicka, {S}peech {R}eplay {D}etection with x-{V}ector
  {A}ttack {E}mbeddings and {S}pectral {F}eatures, in: Proc. INTERSPEECH, 2019.

\bibitem{patel_IS2015}
T.~B~Patel, H.~A. Patil, Combining evidences from mel cepstral, cochlear filter
  cepstral and instantaneous frequency features for detection of natural vs.
  spoofed speech, in: Proc. INTERSPEECH, 2015, pp. 2062--2066.

\bibitem{hannah_raw_spoofing_detection}
H.~{Muckenhirn}, M.~{Magimai-Doss}, S.~{Marcel}, End-to-end convolutional
  neural network-based voice presentation attack detection, in: IEEE
  International Joint Conference on Biometrics (IJCB), 2017, pp. 335--341.

\bibitem{dinkel2017}
H.~Dinkel, N.~Chen, Y.~Qian, K.~Yu, End-to-end spoofing detection with raw
  waveform cldnns, in: Proc. ICASSP, 2017, pp. 4860--4864.

\bibitem{zhang_JSTSP2017}
C.~Zhang, C.~Yu, J.~H. Hansen, An {I}nvestigation of {D}eep {L}earning
  {F}rameworks for {S}peaker {V}erification {A}nti-spoofing, IEEE Journal of
  Selected Topics in Signal Processing.

\bibitem{galina_IS2017}
G.~Lavrentyeva, S.~Novoselov, E.~Malykh, A.~Kozlov, K.~Oleg, V.~Shchemelinin,
  Audio {R}eplay {A}ttack {D}etection with {D}eep {L}earning {F}rameworks, in:
  Proc. INTERSPEECH, 2017, pp. 82--86.

\bibitem{mfccReference}
S.~Davis, P.~Mermelstein, Comparison of parametric representations for
  monosyllabic word recognition in continuously spoken sentences, IEEE
  Transactions on Acoustics, Speech, and Signal Processing 28~(4) (1980)
  357--366.

\bibitem{secondbestsystem_2017challenge}
P.~Nagarsheth, E.~Khoury, K.~Patil, M.~Garland, Replay {A}ttack {D}etection
  {U}sing {DNN} for {C}hannel {D}iscrimination, in: Proc. INTERSPEECH, 2017,
  pp. 97--101.

\bibitem{pca_Jolliffe}
I.~Jolliffe, J.~Cadima, Principal component analysis: A review and recent
  developments, Philosophical Transactions of the Royal Society A:
  Mathematical, Physical and Engineering Sciences 374 (2016) 20150202.
\newblock \href {http://dx.doi.org/10.1098/rsta.2015.0202}
  {\path{doi:10.1098/rsta.2015.0202}}.

\bibitem{lda_tharwat}
A.~Tharwat, T.~Gaber, A.~Ibrahim, A.~E. Hassanien, Linear discriminant
  analysis: {A} detailed tutorial, Ai Communications 30 (2017) 169--190,.
\newblock \href {http://dx.doi.org/10.3233/AIC-170729}
  {\path{doi:10.3233/AIC-170729}}.

\bibitem{goodfellow_GANs}
I.~Goodfellow, J.~Pouget-Abadie, B.~X. M.~Mirza, D.~Warde-Farley, S.~Ozair,
  A.~Courville, Y.~Bengio, Generative adversarial nets, in: Advances in neural
  information processing sys- tems, 2014, pp. 2672--2680.

\bibitem{kingma2013autoencoding}
D.~P. Kingma, M.~Welling, Auto-encoding variational bayes (2013).
\newblock \href {http://arxiv.org/abs/1312.6114} {\path{arXiv:1312.6114}}.

\bibitem{wavenet}
A.~van~den Oord, S.~Dieleman, H.~Zen, K.~Simonyan, O.~Vinyals, A.~Graves,
  N.~Kalchbrenner, A.~Senior, K.~Kavukcuoglu, Wavenet: A generative model for
  raw audio (2016).
\newblock \href {http://arxiv.org/abs/1609.03499} {\path{arXiv:1609.03499}}.

\bibitem{vae_images_label_caption}
Y.~Pu, Z.~Gan, R.~Henao, X.~Yuan, C.~Li, A.~Stevens, L.~Carin, Variational
  autoencoder for deep learning of images, labels and captions, in: Advances in
  neural information processing sys- tems, 2016, p. 2352–2360.

\bibitem{gulrajani2016pixelvae}
I.~Gulrajani, K.~Kumar, F.~Ahmed, A.~A. Taiga, F.~Visin, D.~Vazquez,
  A.~Courville, Pixelvae: A latent variable model for natural images (2016).
\newblock \href {http://arxiv.org/abs/1611.05013} {\path{arXiv:1611.05013}}.

\bibitem{vae_image_forecasting}
J.~Walker, C.~Doersch, A.~Gupta, M.~Hebert, An uncertain future: Forecasting
  from static images using variational autoencoders (2016).
\newblock \href {http://arxiv.org/abs/1606.07873} {\path{arXiv:1606.07873}}.

\bibitem{mocogan}
S.~Tulyakov, M.-Y. Liu, X.~Yang, J.~Kautz, Mocogan: Decomposing motion and
  content for video generation (2017).
\newblock \href {http://arxiv.org/abs/1707.04993} {\path{arXiv:1707.04993}}.

\bibitem{gans_NLP}
S.~Subramanian, S.~Rajeswar, F.~Dutil, C.~Pal, A.~Courville, Adversarial
  generation of natural language, in: Proceedings of the $2^{nd}$ Workshop on
  Representation Learning for {NLP}, Association for Computational Linguistics,
  Vancouver, Canada, 2017, pp. 241--251.
\newblock \href {http://dx.doi.org/10.18653/v1/W17-2629}
  {\path{doi:10.18653/v1/W17-2629}}.

\bibitem{vae_speechModeling}
M.~Blaauw, J.~Bonada, Modeling and {T}ransforming {S}peech using {V}ariational
  {A}utoencoders, in: Proc. INTERSPEECH, 2016, pp. 1770--1774.

\bibitem{hsu_speech_vae_NIPS}
W.~N. Hsu, Y.~Zhang, J.~Glass, Unsupervised learning of disentangled and
  interpretable representations from sequential data, in: Advances in Neural
  Information Processing Systems, 2017.

\bibitem{hsu_speech_vae_Interspeech}
W.~N. Hsu, Y.~Zhang, J.~Glass, Learning latent representations for speech
  generation and transformation, in: Proc. INTERSPEECH, 2017, pp. 1273--1277.

\bibitem{vae_music_synthesis}
P.~Esling, A.~Chemla–Romeu-Santos, A.~Bitton, Generative timbre spaces with
  variational audio synthesis, in: Proc. of the 21st International Conference
  on Digital Audio Effects, 2018.

\bibitem{simon_speech_enh_ICASSP2019}
S.~Leglaive, U.~Simsekli, A.~Liutkus, L.~Girin, R.~Horaud, {Speech enhancement
  with variational autoencoders and alpha-stable distributions}, in: Proc.
  ICASSP, {IEEE}, Brighton, United Kingdom, 2019, pp. 541--545.

\bibitem{vae_multichannel_vc_dataset}
H.~Kameoka, L.~Li, S.~Inoue, S.~Makino, Semi-blind source separation with
  multichannel variational autoencoder (2018).
\newblock \href {http://arxiv.org/abs/1808.00892} {\path{arXiv:1808.00892}}.

\bibitem{feature_learning_asr_vae}
S.~{Tan}, K.~C. {Sim}, Learning utterance-level normalisation using
  {V}ariational {A}utoencoders for robust automatic speech recognition, in:
  IEEE Spoken Language Technology Workshop (SLT), 2016, pp. 43--49.

\bibitem{asr_applications_vae}
S.~Feng, T.~Lee, {I}mproving {U}nsupervised {S}ubword {M}odeling via
  {D}isentangled {S}peech {R}epresentation {L}earning and {T}ransformation, in:
  Proc. INTERSPEECH, 2019.

\bibitem{dataaugmentation_vae_asv}
Z.~Wu, S.~Wang, Y.~Qian, K.~Yu, {D}ata {A}ugmentation using {V}ariational
  {A}utoencoder for {E}mbedding based {S}peaker {V}erification, in: Proc.
  INTERSPEECH, 2019.

\bibitem{regularisation_vae_asv}
Y.~Zhang, L.~Li, D.~Wang, {VAE}-based regularization for deep speaker
  embedding, in: Proc. INTERSPEECH, 2019.

\bibitem{adaptation_vae_asv}
Y.~Tu, M.~W. Mak, J.~T. Chien, {V}ariational {D}omain {A}dversarial {L}earning
  for {S}peaker {V}erification, in: Proc. INTERSPEECH, 2019.

\bibitem{zhang_vae_tts}
Y.~{Zhang}, S.~{Pan}, L.~{He}, Z.~{Ling}, Learning latent representations for
  style control and transfer in end-to-end speech synthesis, in: Proc. ICASSP,
  2019, pp. 6945--6949.

\bibitem{yexin2019challenge}
Y.~Yang, H.~Wang, H.~Dinkel, Z.~Chen, S.~Wang, Y.~Qian, K.~Yu, {T}he {SJTU}
  {R}obust {A}nti-spoofing {S}ystem for the {ASV}spoof 2019 {C}hallenge, in:
  Proc. INTERSPEECH, 2019.

\bibitem{Huang_vae_gan}
H.~Huang, Z.~Li, R.~He, Z.~Sun, T.~Tan, Intro{VAE}: {I}ntrospective
  {V}ariational {A}utoencoders for {P}hotographic {I}mage {S}ynthesis, in:
  Proc. of the $32^{nd}$ International Conference on Neural Information
  Processing Systems, NIPS'18, Curran Associates Inc., USA, 2018, pp. 52--63.

\bibitem{conditional_vae}
K.~Sohn, H.~Lee, X.~Yan, Learning structured output repre- sentation using deep
  conditional generative models, in: Advances in Neural Information Processing
  Systems, 2015, p. 3483–3491.

\bibitem{speech_separation_vae_auxiliary_classifier_ICASSP}
L.~{Li}, H.~{Kameoka}, S.~{Makino}, Fast {MVAE}: {J}oint {S}eparation and
  {C}lassification of {M}ixed {S}ources {B}ased on {M}ultichannel {V}ariational
  {A}utoencoder with {A}uxiliary {C}lassifier, in: Proc. ICASSP, 2019, pp.
  546--550.

\bibitem{auxiliary_classifier_vae_vc}
H.~Kameoka, T.~Kaneko, K.~Tanaka, N.~Hojo, {ACVAE}-{VC}: {N}on-parallel
  many-to-many voice conversion with auxiliary classifier variational
  autoencoder (2018).
\newblock \href {http://arxiv.org/abs/1808.05092} {\path{arXiv:1808.05092}}.

\bibitem{bhusanSLT2018}
B.~Chettri, S.~Mishra, B.~L. Sturm, E.~Benetos, Analysing the {P}redictions of
  a {CNN}-based {R}eplay {S}poofing {D}etection {S}ystem, in: IEEE
  International Workshop on Spoken Language Technology (SLT), 2018.

\bibitem{wu2015light}
X.~Wu, R.~He, Z.~Sun, T.~Tan, A light cnn for deep face representation with
  noisy labels (2015).
\newblock \href {http://arxiv.org/abs/1511.02683} {\path{arXiv:1511.02683}}.

\bibitem{yanmin2016}
Y.~Qian, N.~Chen, K.~Yu, {Deep features for automatic spoofing detection},
  Speech Communication.

\bibitem{kaavya_IS2018}
K.~Sriskandaraja, V.~Sethu, E.~Ambikairajah, Deep {S}iamese {A}rchitecture
  {B}ased {R}eplay {D}etection for {S}ecure {V}oice {B}iometric, in: Proc.
  INTERSPEECH, 2018.

\bibitem{bayesian2019challenge}
R.~Bialobrzeski, M.~Kosmiderm, M.~Matuszewski, M.~Plata, A.~Rakowski, {R}obust
  {B}ayesian and {L}ight {N}eural {N}etworks for {V}oice {S}poofing
  {D}etection, in: Proc. INTERSPEECH, 2019.

\bibitem{but2019challenge}
H.~Zeinali, T.~Stafylakis, J.~R. Georgia~Athanasopoulou, I.~Gkinis, L.~Burget,
  J.~H. Cernocky, Detecting {S}poofing {A}ttacks {U}sing {VGG} and {S}inc{N}et:
  {BUT}-{O}milia {S}ubmission to {ASV}spoof 2019 {C}hallenge, in: Proc.
  INTERSPEECH, 2019.

\bibitem{alejandro2019challenge}
A.~Gomez-Alanis, A.~M. Peinado, J.~A. Gonzalez, A.~M. Gomez, {A} {L}ight
  {C}onvolutional {GRU}-{RNN} {D}eep {F}eature {E}xtractor for {ASV} {S}poofing
  {D}etection, in: Proc. INTERSPEECH, 2019.

\bibitem{chang2019challenge}
S.-Y. Chang, K.-C. Wu, C.-P. Chen, {}transfer-{R}epresentation {L}earning for
  {D}etecting {S}poofing {A}ttacks with {C}onverted and {S}ynthesized {S}peech
  in {A}utomatic {S}peaker {V}erification {S}ystem, in: Proc. INTERSPEECH,
  2019.

\bibitem{he2015deep}
K.~He, X.~Zhang, S.~Ren, J.~Sun, Deep {R}esidual {L}earning for {I}mage
  {R}ecognition (2015).
\newblock \href {http://arxiv.org/abs/1512.03385} {\path{arXiv:1512.03385}}.

\bibitem{weicheng2019challenge}
W.~Cai, H.~Wu, D.~Cai, M.~Li, {T}he {DKU} {R}eplay {D}etection {S}ystem for the
  {ASV}spoof 2019 {C}hallenge: {O}n {D}ata {A}ugmentation, {F}eature
  {R}epresentation, {C}lassification, and {F}usion, in: Proc. INTERSPEECH,
  2019.

\bibitem{rongjin2019challenge}
R.~Li, M.~Zhao, Z.~Li, L.~Li, Q.~Hong, {A}nti-{S}poofing {S}peaker
  {V}erification {S}ystem with {M}ulti-{F}eature {I}ntegration and
  {M}ulti-{T}ask {L}earning, in: Proc. INTERSPEECH, 2019.

\bibitem{fourthbestsystem_2017challenge}
Z.~Chen, Z.~Xie, W.~Zhang, X.~Xu, Res{N}et and {M}odel {F}usion for {A}utomatic
  {S}poofing {D}etection, in: Proc. INTERSPEECH, 2017, pp. 102--106.

\bibitem{assert2019challenge}
C.-I. Lai, N.~Chen, J.~Villalba, N.~Dehak, {ASSERT}: {A}nti-{S}poofing with
  {S}queeze-{E}xcitation and {R}esidual ne{T}works, in: Proc. INTERPSEECH,
  2019.

\bibitem{moustafa2019challenge}
M.~Alzantot, Z.~Wang, M.~B. Srivastava, {D}eep {R}esidual {N}eural {N}etworks
  for {A}udio {S}poofing {D}etection, in: Proc. INTERSPEECH, 2019.

\bibitem{jee2019challenge}
J.~weon Jung, H.~jin Shim, H.-S. Heo, H.-J. Yu, {R}eplay attack detection with
  complementary high-resolution information using end-to-end {DNN} for the
  {ASV}spoof 2019 {C}hallenge, in: Proc. INTERSPEECH, 2019.

\bibitem{resnet_dataAugmentation}
W.~Cai, C.~Danwei, W.~Liu, G.~Li, M.~Li, Countermeasures for {A}utomatic
  {S}peaker {V}erification {R}eplay {S}poofing {A}ttack : {O}n {D}ata
  {A}ugmentation, {F}eature {R}epresentation, {C}lassification and {F}usion,
  in: Proc. INTERSPEECH, 2017, pp. 17--21.

\bibitem{liuICASSP2019}
M.~{Liu}, L.~{Wang}, J.~{Dang}, S.~{Nakagawa}, H.~{Guan}, X.~{Li}, Replay
  {A}ttack {D}etection {U}sing {M}agnitude and {P}hase {I}nformation with
  {A}ttention-based {A}daptive {F}ilters, in: Proc. ICASSP, 2019, pp.
  6201--6205.

\bibitem{laiICASSP2019}
C.~{Lai}, A.~{Abad}, K.~{Richmond}, J.~{Yamagishi}, N.~{Dehak}, S.~{King},
  Attentive {F}iltering {N}etworks for {A}udio {R}eplay {A}ttack {D}etection,
  in: Proc. ICASSP, 2019, pp. 6316--6320.

\bibitem{wu_IS2015}
Z.~Wu, T.~Kinnunen, N.~Evans, J.~Yamagishi, C.~Hanilci, M.~Sahidullah,
  A.~Sizov, {ASV}spoof 2015: the {F}irst {A}utomatic {S}peaker {V}erification
  {S}poofing and {C}ountermeasures {C}hallenge, in: Proc. INTERSPEECH, 2015.

\bibitem{tomiSummaryPaper}
T.~Kinnunen, M.~Sahidullah, H.~Delgado, M.~Todisco, N.~Evans, J.~Yamagishi,
  K.~A. Lee, The {ASV}spoof 2017 {C}hallenge: {A}ssessing the {L}imits of
  {R}eplay {S}poofing {A}ttack {D}etection, in: Proc. INTERSPEECH, 2017.

\bibitem{asvspoof2019overview}
M.~Todisco, X.~Wang, V.~Vestman, M.~Sahidullah, H.~Delgado, A.~Nautsh,
  J.~Yamagishi, N.~Evans, T.~Kinnunen, K.~A. Lee, {ASV}spoof 2019: {F}uture
  {H}orizons in {S}poofed and {F}ake {A}udio {D}etection, in: Proc.INTERSPEECH,
  2019.

\bibitem{hector_cqcc}
M.~Todisco, H.~Delgado, N.~Evans, Constant {Q} cepstral coefficients: A
  spoofing countermeasure for automatic speaker verification, Computer Speech
  and Language, Volume 45, (2017) Pages 516--535.

\bibitem{rohan_eCQCC}
J.~{Yang}, R.~K. {Das}, H.~{Li}, Extended {C}onstant-{Q} {C}epstral
  {C}oefficients for {D}etection of {S}poofing {A}ttacks, in: 2018 Asia-Pacific
  Signal and Information Processing Association Annual Summit and Conference
  (APSIPA ASC), 2018, pp. 1024--1029.

\bibitem{patel_IS2017}
H.~A. Patil, M.~R. Kamble, T.~B. Patel, M.~Soni, {N}ovel {V}ariable {L}ength
  {T}eager {E}nergy {S}eparation {B}ased {I}nstantaneous {F}requency {F}eatures
  for {R}eplay {D}etection, in: Proc. INTERSPEECH, 2017.

\bibitem{madhu_IS2018}
M.~R.~Kamble, H.~Tak, H.~A.~Patil, Effectiveness of {S}peech
  {D}emodulation-{B}ased {F}eatures for {R}eplay {D}etection, in: Proc.
  INTERSPEECH, 2018.

\bibitem{buddhi_IS2018}
B.~Wickramasinghe, S.~Irtza, E.~Ambikairajah, J.~Epps, Frequency domain linear
  prediction features for replay spoofing attack detection, in: Proc.
  INTERSPEECH, 2018.

\bibitem{tharshini_IS2018}
T.~Gunendradasan, B.~Wickramasinghe, P.~Ngoc~Le, E.~Ambikairajah, J.~Epps,
  Detection of {R}eplay-{S}poofing {A}ttacks using {F}requency {M}odulation
  {F}eatures, in: Proc. INTERSPEECH, 2018.

\bibitem{saranya_IS2018}
S.~M. S, H.~A. Murthy, Decision-level feature switching as a paradigm for
  replay attack detection, in: Proc. INTERSPEECH, 2018.

\bibitem{buddhi2019IS}
B.~Wickramasinghe, E.~Ambikairajah, J.~Epps, {B}iologically {I}nspired
  {A}daptive-{Q} {F}ilterbanks for {R}eplay {S}poofing {A}ttack {D}etection,
  in: Proc. INTERSPEECH, 2019.

\bibitem{hardik_IS2018}
H.~B.~Sailor, M.~R.~Kamble, H.~A.~Patil, Auditory {F}ilterbank {L}earning for
  {T}emporal {M}odulation {F}eatures in {R}eplay {S}poof {S}peech {D}etection,
  in: Proc. INTERSPEECH, 2018.

\bibitem{ravanelli2018speaker}
M.~Ravanelli, Y.~Bengio, Speaker {R}ecognition from {R}aw {W}aveform with
  {S}inc{N}et (2018).
\newblock \href {http://arxiv.org/abs/1808.00158} {\path{arXiv:1808.00158}}.

\bibitem{bhusan2019challenge}
B.~Chettri, D.~Stoller, V.~Morfi, M.~A.~M. Ram\'{i}rez, E.~Benetos, B.~L.
  Sturm, {E}nsemble {M}odels for {S}poofing {D}etection in {A}utomatic
  {S}peaker {V}erification, in: Proc. INTERSPEECH, 2019, pp. 1018--1022.

\bibitem{Altosaar2019-vae-tutorial}
J.~Altosaar,
  \href{https://jaan.io/what-is-variational-autoencoder-vae-tutorial/}{Tutorial
  -- what is a variational autoencoder?} (2019).
\newline\urlprefix\url{https://jaan.io/what-is-variational-autoencoder-vae-tutorial/}

\bibitem{Doersch16-VAE-tutorial}
C.~Doersch, \href{http://arxiv.org/abs/1606.05908}{Tutorial on variational
  autoencoders}, CoRR abs/1606.05908.
\newblock \href {http://arxiv.org/abs/1606.05908} {\path{arXiv:1606.05908}}.
\newline\urlprefix\url{http://arxiv.org/abs/1606.05908}

\bibitem{bishop_2006}
C.~M. Bishop, Pattern recognition and machine learning, Vol.~1, springer, New
  York, 2006.

\bibitem{CoverThomas2001-elements}
T.~M. Cover, J.~A. Thomas, \href{https://doi.org/10.1002/0471200611}{Elements
  of Information Theory}, Wiley, 2001.
\newblock \href {http://dx.doi.org/10.1002/0471200611}
  {\path{doi:10.1002/0471200611}}.
\newline\urlprefix\url{https://doi.org/10.1002/0471200611}

\bibitem{cvae_av_speech_synthesis}
S.~Dahmani, V.~Colotte, V.~Girard, S.~Ouni, {C}onditional {V}ariational
  {A}uto-{E}ncoder for {T}ext-{D}riven {E}xpressive {A}udio{V}isual {S}peech
  {S}ynthesis, in: Proc. INTERSPEECH, 2019.

\bibitem{Dempster1977-EM}
A.~P. Dempster, N.~M. Laird, D.~B. Rubin,
  \href{http://web.mit.edu/6.435/www/Dempster77.pdf}{Maximum likelihood from
  incomplete data via the {EM} algorithm}, Journal of the Royal Statistical
  Society: Series B 39 (1977) 1--38.
\newline\urlprefix\url{http://web.mit.edu/6.435/www/Dempster77.pdf}

\bibitem{Jin2016-local-maxima-GMM}
C.~Jin, Y.~Zhang, S.~Balakrishnan, M.~J. Wainwright, M.~I. Jordan, Local maxima
  in the likelihood of gaussian mixture models: Structural results and
  algorithmic consequences, in: Advances in Neural Information Processing
  Systems 29: Annual Conference on Neural Information Processing Systems 2016,
  December 5-10, 2016, Barcelona, Spain, 2016, pp. 4116--4124.

\bibitem{hectorAsvspoof2.0}
H.~Delgado, M.~Todisco, M.~Sahidullah, N.~Evans, T.~Kinnunen, K.~Lee,
  J.~Yamagishi, {ASVspoof 2017 Version 2.0: meta-data analysis and baseline
  enhancements}, in: Proc. Speaker Odyssey, 2018.

\bibitem{redDotsDataCollection}
K.~A. Lee, A.~Larcher, G.~Wang, P.~Kenny, N.~Brummer, D.~van Leeuwen,
  H.~Aronowitz, M.~Kockmann, C.~Vaquero, B.~Ma, H.~Li, T.~Stafylakis, J.~Alam,
  A.~Swart, J.~Perez., The {R}ed{D}ots {D}ata {C}ollection for {S}peaker
  {R}ecognition,, in: Proc. INTERSPEECH, 2015.

\bibitem{kinnunen2017reddots}
T.~Kinnunen, M.~Sahidullah, M.~Falcone, L.~Costantini, R.~G. Hautamaki,
  D.~A.~L. Thomsen, A.~K. Sarkar, Z.-H. Tan, H.~Delgado, M.~Todisco, et~al.,
  {R}eddots {R}eplayed: {A} {N}ew {R}eplay {S}poofing {A}ttack {C}orpus {F}or
  {T}ext-{D}ependent {S}peaker {V}erification {R}esearch, in: Proc. ICASSP,
  2017.

\bibitem{asvspoof2019_evaluationplan}
{ASV}spoof 2019,
  \href{http://www.asvspoof.org/asvspoof2019/asvspoof2019_evaluation_plan.pdf}{the
  {A}utomatic {S}peaker {V}erification {S}poofing and {C}ountermeasures
  {C}hallenge {E}valuation {P}lan.}
\newline\urlprefix\url{http://www.asvspoof.org/asvspoof2019/asvspoof2019_evaluation_plan.pdf}

\bibitem{mishra2019ganbased}
S.~Mishra, D.~Stoller, E.~Benetos, B.~L. Sturm, S.~Dixon, Gan-based
  {G}eneration and {A}utomatic {S}election of {E}xplanations for {N}eural
  {N}etworks (2019).
\newblock \href {http://arxiv.org/abs/1904.09533} {\path{arXiv:1904.09533}}.

\bibitem{dumoulin2016guide}
V.~Dumoulin, F.~Visin, A guide to convolution arithmetic for deep learning
  (2016).
\newblock \href {http://arxiv.org/abs/1603.07285} {\path{arXiv:1603.07285}}.

\bibitem{leakyRelu}
A.~L. Maas, A.~Y. Hannun, A.~Y. Ng, Rectifier nonlinearities improve neural
  network acoustic models, ICML, 2013.

\bibitem{adam}
D.~P. Kingma, J.~Ba, \href{http://arxiv.org/abs/1412.6980}{Adam: {A} method for
  stochastic optimization}, CoRR abs/1412.6980.
\newblock \href {http://arxiv.org/abs/1412.6980} {\path{arXiv:1412.6980}}.
\newline\urlprefix\url{http://arxiv.org/abs/1412.6980}

\bibitem{tomi_tDCF}
T.~Kinnunen, K.~Lee, H.~Delgado, N.~Evans, M.~Todisco, M.~Sahidullah,
  J.~Yamagishi, D.A, Reynolds, {t-DCF: a Detection Cost Function for the Tandem
  Assessment of Spoofing Countermeasures and Automatic Speaker Verification},
  in: Proc. Speaker Odyssey, 2018.

\bibitem{tsne_Maaten2008VisualizingDU}
L.~van~der Maaten, G.~E. Hinton, {V}isualizing {D}ata using t-{SNE}, Journal of
  Machine Learning Research 1 (2008) 1--48.

\end{thebibliography}

\end{document}